\documentclass[twocolumn]{aastex63}
\shorttitle{Free-form lens model of El Gordo}
\shortauthors{Diego et al.}
\usepackage{txfonts}
\usepackage[T1]{fontenc}
\newcommand{\vectX}{\bf {\it X}}
\newcommand{\vectTheta}{\bf {\it \Theta}}
\newcommand{\matrGamma}{\bf \Gamma}
\begin{document}

\title{Free-form lens model and mass estimation of the high redshift galaxy cluster ACT-CL J0102-4915, ``El Gordo''}

\correspondingauthor{Jose M. Diego}
\email{jdiego@ifca.unican.es}

\author[0000-0001-9065-3926]{Jose M. Diego}
\affiliation{Instituto de F\'isica de Cantabria (CSIC-UC). Avda. Los Castros s/n. 39005 Santander, Spain}  %1

\author{S.M. Molnar}
\affiliation{Institute of Astronomy and Astrophysics, Academia Sinica, P.O. Box 23-141, Taipei 10617, Taiwan} %2

\author{C. Cerny}
\affiliation{Department of Astronomy, University of Michigan, 1085 South University Avenue, Ann Arbor, MI 48109} %3

\author{T. Broadhurst}
\affiliation{Department of Theoretical Physics, University of the Basque Country, Bilbao E-48080, Spain} %4 
\affiliation{Ikerbasque, Basque Foundation for Science, Alameda Urquijo, 36-5 Plaza Bizkaia E-48011, Bilbao, Spain} %5

\author{R. Windhorst}
\affiliation{School of Earth and Space Exploration, Arizona State University, Tempe, AZ 85287-1404, USA} %6

\author{A. Zitrin}
\affiliation{Department of Physics, Ben-Gurion University, Be'er-Sheva 84105, Israel} %7

\author{R. Bouwens}
\affiliation{Leiden Observatory, Leiden University, NL-2300 RA Leiden, The Netherlands} %8

\author{D. Coe}
\affiliation{Space Telescope Science Institute, Baltimore, MD, USA} %9

\author{C. Conselice}
\affiliation{Centre for Astronomy and Particle Theory, School of Physics \& Astronomy, University of Nottingham, Nottingham, NG7 2RD, UK} % 10

\author{K. Sharon}
\affiliation{Department of Astronomy, University of Michigan, 1085 South University Avenue, Ann Arbor, MI 48109} %3
%\nocollaboration{1}




%   \date{Received September 15, 1996; accepted March 16, 1997}

 \begin{abstract}
We examine the massive colliding cluster El Gordo, one of the most massive clusters at high redshift. We use a free-form lensing reconstruction method that avoids making assumptions about the mass distribution. We use data from the RELICS program and identify new multiply lensed system candidates. The new set of constraints and free-form method provides a new independent mass estimate of this intriguing colliding cluster. Our results are found to be consistent with earlier parametric models, indirectly confirming the assumptions made in earlier work. By fitting a double gNFW profile to the lens model, and extrapolating to the virial radius, we infer a total mass for the cluster of $M_{200c}=(1.08^{+0.65}_{-0.12})\times10^{15}$M$_{\odot}$.  We estimate the uncertainty in the mass due to errors in the photometric redshifts, and discuss the uncertainty in the inferred virial mass due to the extrapolation from the lens model. We also find in our lens map a mass overdensity corresponding to the large cometary tail of hot gas, reinforcing its interpretation as a large tidal feature predicted by hydrodynamical simulations that mimic El Gordo. Finally, we discuss the observed relation between the plasma and the mass map, finding that the peak in the projected mass map may be associated with a large concentration of colder gas, exhibiting possible star formation. El Gordo is one of the first clusters that will be observed with JWST, which is expected to unveil new high redshift lensed galaxies around this interesting cluster, and provide a more accurate estimation of its mass.

 \end{abstract}

%   \keywords{gravitational lensing --
%                microlensing -- dark matter -- 
%                cosmology
%               }
%
%-------------------------------------------------------------------

\section{Introduction}
%%%%%%%%%%%%%%%%%%%%%%%%
The galaxy cluster, ACT-CL J0102-4915, also known as {\it El Gordo}, is a relatively high redshift z=0.870 with a rich, bimodal galaxy distribution \citep{Williamson2011,Menanteau2012}. Its mass has been estimated in earlier work using different techniques, including combined dynamical, X-ray and Sunyaev-Zeldovich (SZ) data \citep{Menanteau2012}, strong lensing data \citep{Zitrin2013,Cerny2018}, and weak lensing data \citep{Jee2014}. El Gordo has also been the subject of several dynamical studies, including numerical N-body simulations \citep{Donnert2014} and hydrodynamical simulations \citep{Molnar2015,Zhang2015,Zhang2018}. These studies have highlighted the impressively large scale cometary structure visible clearly in X-ray images, that appears to imply El Gordo is being observed right after a collision of two subgroups \citep{Molnar2015,Zhang2015}, similar to the iconic {\it Bullet} cluster. This interpretation is supported by the presence of two radio relics ahead and behind the X-ray cometary structure \citep{Molnar2015,Lindner2014}. Based on the X-ray and radio morphology, as well as on a preliminary lens model for the mass distribution, \cite{Ng2015} argue that El Gordo is in a return phase after first core passage. This means that the cluster is being observed after the phase of maximum apocenter, and that the two groups are moving against each other rather than away from each other. Part of this conclusion is based on a lens model that relies on weak lensing and that assigns more mass to the NW clump than to the SE clump. This interpretation is however challenged by lens models based on strong lensing data that place most of the mass in the SE group \citep{Zitrin2013,Cerny2018}. 

The El Gordo cluster is an extreme cluster at several levels. It is the most massive cluster at $z\approx 0.9$ with an estimated mass ranging from $M_{200c} = 1.8\times 10^{15}M_{\odot}$ to $M_{200c} = 2 \times 10^{15}M_{\odot}$. $M_{200c}$ is the mass within the sphere of radius $r_{200}$. This radius is defined as the radius where the mass density enclosed in the sphere with the same radius, and centered in the object is 200 times the critical density of the universe at the cluster redshift. Some authors estimate the overdensity in relation to the average density of the universe, $\rho_m = \rho_c\Omega_m$. In this case the mass is denoted as $M_{200\rho_m}$.
Using SZ data, \cite{Williamson2011} estimates a mass of $M_{200\rho_m} = 1.89 \pm 0.45\times 10^{15}M_{\odot}$. \cite{Menanteau2012} obtained a mass estimate of $M_{200\rho_m} = 2.16 \pm 0.32 \times 10^{15}M_{\odot}$ which relied on different scaling relations. Based on an extrapolation of the strong lensing mass model, \cite{Zitrin2013} estimate a total mass of  $M_{200\rho_m} \sim 2.3 \times 10^{15}M_{\odot}$.  Using weak lensing obtained with HST,  \cite{Jee2014} estimates $M_{200c} = 3.13 \pm 0.56 \times 10^{15}M_{\odot}$. 

At the high-end of the mass range for El Gordo, these masses are in tension with the standard LCDM model \citep[see for instance][]{Jee2014}, that predicts the maximum mass at this redshift should be less than $M_{200\rho_m} \approx 1.7 \times 10^{15}M_{\odot}$ \citep{Harrison2012}. A similar conclusion is reached from large N-body simulations. Using the very large 630 Gpc$^3$ N-body simulation Jubilee (based on a standard LCDM model), \cite{Watson2014} find that the most massive cluster in the simulation and at $z=0.9$ is $M_{200\rho_m} \approx 1.5\times 10^{15}M_{\odot}$ (see their Figure 5). Note that in \cite{Watson2014}, the masses are defined as $M_{178\rho_m}$ rather than $M_{200\rho_m}$ or $M_{200c}$. For an NFW profile, $M_{178\rho_m} \approx 1.2M_{200c}$ and $M_{178\rho_m}\approx 4\%$ times higher than $M_{200\rho_m}$ \citep{Waizmann2012}. 
%\footnote{Note that in this reference, the masses are defined as $M_{178\rho_m}$ rather than $M_{200\rho_m}$ or $M_{200c}$. For an NFW profile, $M_{178\rho_m} \approx 1.7M_{200c}$ and $M_{178\rho_m}\approx 4\%$ times higher than $M_{200\rho_m}$}. 
Given the fact that El Gordo was found in a relatively small area of the sky (the footprint of the original ACT survey covers less than 2\% the area of the sky), it raises the question about its significance as possible evidence for tension with the LCDM model. How likely is to find such an extreme object in an area smaller than 2\% of the sky? The tension could be reduced if previous mass estimates are found to be too high. Additional mass estimates, based on alternative methods, can explore the uncertainty in the mass of the cluster. In this paper we present such an alternative estimate based on the first free-form lensing modeling of this cluster. Free-form models are affected by different systematics than parametric models. Hence, it is important to estimate the mass of extreme clusters (like El Gordo) with as many independent estimators as possible. We note, that after correcting for redshift and Eddington bias, \cite{Waizmann2012} finds El Gordo not to be in tension with LCDM. However, in their analysis the survey area is assumed to be 3.7 times higher than the area where El Gordo was originally found.
%At the source of this apparent tension could be an overestimation of its mass. Hence, it is important to improve on its mass estimation using the latest lensing data, as well as alternative lensing models that rely on the same lensing data, so possibly systematic effects inherent to the lensing reconstruction technique employed can be addressed.
 
As of 2014, El Gordo was also the highest-redshift cluster known to host radio relics \citep{Lindner2014}. The X-ray emission exhibits an interesting offset between the peak of the X-ray emission and the position of the BCG. Contrary to what happens in the Bullet cluster, the X-ray peak seems to be ahead of the BCG. However, in the interpretation of \cite{Ng2015}, the BCG would be moving towards the second group, so the X-ray peak would be trailing the BCG. The returning phase interpretation of El Gordo is challenged based on results from dedicated N-body/hydrodynamical simulations reproducing most of the observations of El Gordo \citep{Molnar2015,Zhang2015}.
In \cite{Molnar2018}, the authors demonstrate that the speed of the outgoing shocks can be very large (4000--5000 km s$^{-1}$) in a massive merging cluster like El Gordo, therefore leaving the system before the first turnaround.
%FROM Molnar \& Broadhurst (2018): “We have demonstrated, that the outgoing shocks travel fast (4000-5000 km s−1) in the low density outer gas of the two subclusters, and, therefore, their positions relative to those of the mass peaks can be used to accurately derive the phase of the collision. Thus, merging shocks can (re)accelerate particles and generate relics only for a limited time <∼5 × 108 yr, so that shocks and bright relics can be detected in a merging system soon after the first core passage, before the first turnaround. After that time the relics become fainter after no shocks feed them as the electrons loose energy. This point has been unappreciated in earlier work (Golovich et al. 2017; Ng et al. 2015) where later stage merging has been entertained without the guidance of hydrodynamical simulations like those presented in this paper.”

El Gordo is also unique in the sense that it is a powerful lens at relatively high redshift.  One of the features that makes El Gordo an attractive target for lensing studies is the fact that for sources at high redshift, critical curves form at relatively large distances from the member galaxies. This is particularly true in the gap between the two clusters, where the critical curves are relatively undisturbed by nearby member galaxies. Having undisturbed critical curves is relevant to observe caustic crossing events of distant stars \citep{Kelly2018,Diego2018}, since in this case the maximum magnification can be larger than in situations where critical curves are affected by microlenses in member galaxies, or from the intracluster medium. Caustic crossing events has been proposed as a technique useful to study Pop III stars and stellar-mass black hole accretion discs in \cite{Windhorst2018} with JWST. 
Because El Gordo is the highest redshift known cluster with potentially such significant transverse motion --- based on the X-ray
morphology and the two lensing mass centers discussed in this paper --- it is an ideal target for JWST follow-up to search for caustic transits at z$>>$1, and possibly for First Light caustic transits at z$>$7. For this reason, El Gordo is a JWST GTO target that will be observed in Cycle 1 (JWST program \# 1176; PI: Windhorst). It is our sincere hope that JWST Guest Observers will propose to observe El Gordo in many successive epochs, amongst others to find caustic transits at z$>>$1.

In this paper we derive the mass distribution and study this interesting cluster using the latest public data from the RELICS program, and newly identified strong lensing system candidates. We use our free-form lensing reconstruction code WSLAP+ \citep{Diego2005,Diego2007,Sendra2014,Diego2016}, which does not rely on assumptions about the distribution of dark matter. Our results offer an important cross-check with previous results, since any disagreement between our free-form method and results obtained by previous parametric methods could signal potential systematic problems in one (or both) types of modeling. We pay special attention to the integrated mass as a function of radius and the effect that extrapolations of the derived mass profile up to the virial radius has on the inferred total mass of the cluster.   

This paper is organized as follows. In section \ref{sect_data} we describe the data and simulations used in this work. 
In section \ref{sec_math} we describe briefly the algorithm used to perform the lensing reconstruction.  Results are presented 
in section \ref{sect_result} and discussed in section \ref{sect_disc}. We summarize and conclude in section \ref{sect_concl}. 
We adopt a standard flat cosmological model with $\Omega_m=0.3$ and $h=0.7$. At the redshift of the lens, and for this cosmological model, one arcsecond corresponds to 7.8 kpc. 

%----------------------------------------------------------------- 
   \begin{figure*}
%%   \centering
   \plotone{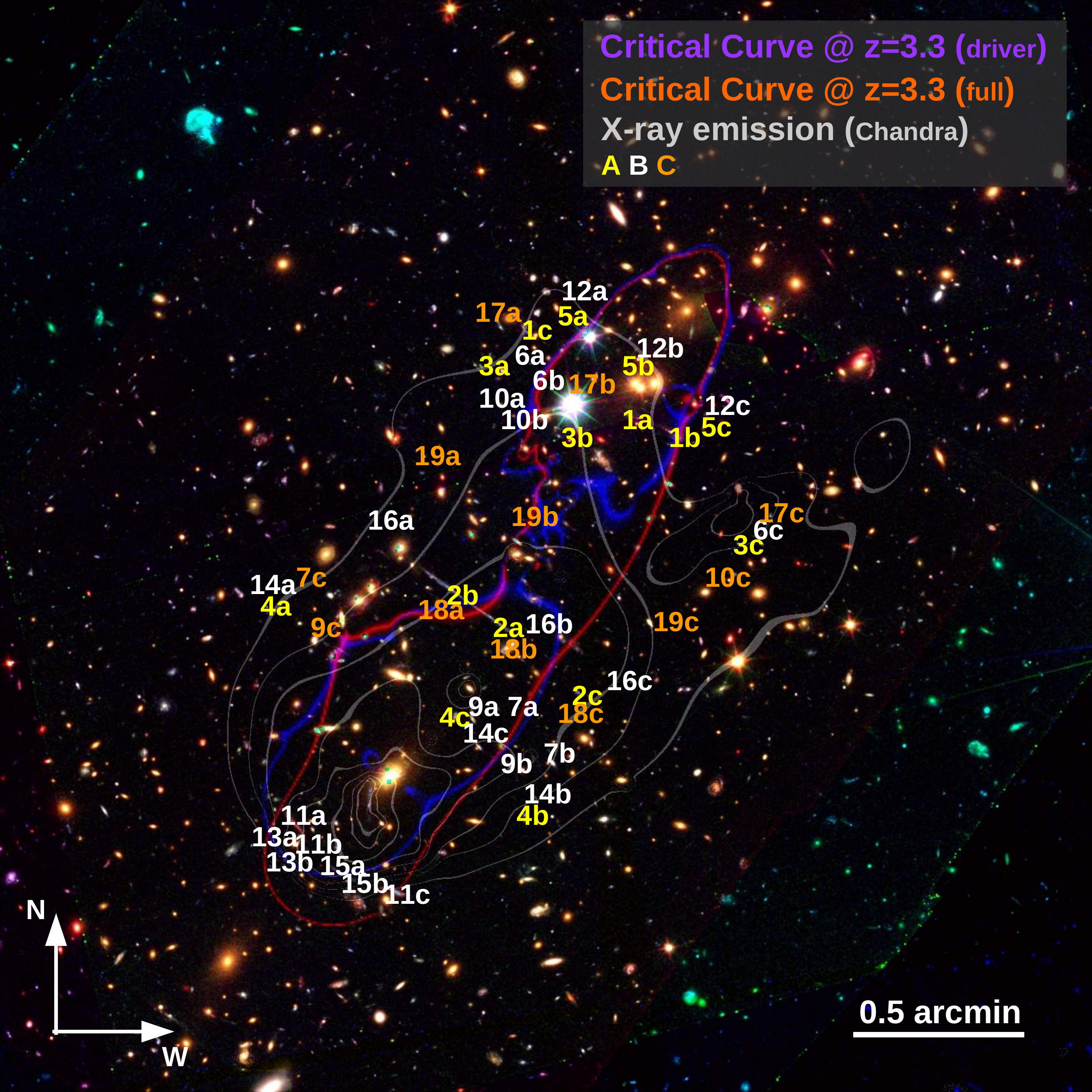}
      \caption{HST Optical+IR composite image with overlaid contours from Chandra. Multiply lensed images are marked with their corresponding IDs. 
               The color of the IDs indicate the quality of the family identification. Images with yellow IDs are category A (reliable), 
               IDs in white are still reliable, but not as confident as A. Images with IDs in orange are less reliable although still valid candidates. 
               The blue and red curves show the critical curve at $z=3.3$ for the driver model (derived from images in category A) 
               and full model (derived from images in categories A and B) respectively. 
              }
         \label{Fig_CC_Arcs_Chandra}
   \end{figure*}
%-----------------------------------------------------------------

\section{Observational and simulated data}\label{sect_data}
%%%%%%%%%%%%%%%%%%%%%%%%%%%%%%%%%%%%%%%%%%%%%%%%%%%%%%%%%%%%
In this section we describe briefly the data used in this work, as well as previous N-body simulations of El Gordo cluster that will be used to compare with our results. 

\subsection{Optical data}
%---------------------------
%We use the reduced public imaging data obtained from the ACS and WFC3 Hubble instruments, retrieved from the Mikulski Archive for Space Telescope (MAST) and from programs ID: GO 12755 P.I J. Hughes, GO 12477 P.I F. High, and GO 14096 P.I D. Coe. This cluster was also recently observed as part of the Reionization Lensing Cluster Survey (RELICS) program \citep{Coe2019}. RELICS observes 41 clusters with typically 4--5 orbits per cluster. Observations are made both in the optical with the ACS camera and in the IR with the WFC3 camera. The frequency coverage ranges between $approx 0.4 \mu$m to $\approx 1.6 \mu$m. From the original files, we produce colour images by combining the optical and IR bands. 
We use public Hubble imaging data from programs GO 12755 (PI J. Hughes), GO 12477 (PI F. High), and GO 14096 (PI D. Coe). These ACS and WFC3/IR observations include data in 10 filters spanning wavelengths $sim$0.4--1.7 $\mu$m. The Reionization Lensing Cluster Survey \citep[RELICS;][]{Coe2019} delivered reduced images combining data from all of these HST programs, including their own (14096).
RELICS also delivered photometric redshift catalogs of objects detected in these images.  
We retrieved these data products from the Mikulski Archive for Space Telescope (MAST).
From the reduced images, we produce colour images by combining the optical and IR bands. A parametric lens model derived using the new RELICS (and previous HST) data is presented in \cite{Cerny2018}.

\subsection{X-ray data}
%---------------------------
%\dataset [ADS/Sa.CXO#obs/12258] {Chandra ObsId 12258}
%\dataset [ADS/Sa.CXO#obs/14022] {Chandra ObsId 14022}
%\dataset [ADS/Sa.CXO#obs/14023] {Chandra ObsId 14023}
To explore the dynamical state of El Gordo, we also produce an X-ray image using public Chandra data. In particular, we used data from the ACIS instrument acquired in 2011--2012 and with the Obs ID 12258, Obs ID  14022 and Obs ID 14023 (PI. J. Hughes) totaling $\approx 350$ ks. The X-ray data is smoothed using the code {\small ASMOOTH} \citep{Ebeling2006}. 
A false color image from the HFF imaging overlaid contours of the smoothed X-ray data is shown in figure \ref{Fig_CC_Arcs_Chandra}.
The distribution of X-rays shows a cometary structure similar to the Bullet cluster. The peak of the X-ray emission is offset with respect to the BCG by $\approx 70$ kpc. 

\subsection{Simulated data}
%---------------------------
In order to study the impact of extrapolating a strong lensing derived profile (i.e, a profile that extends a limited range in radii) up to the virial radius, we use dedicated N-body/hydrodynamical simulations that mimic El Gordo \citep[for details of the simulations see][]{Molnar2015}. Our simulations were constrained by multi-frequency data: X-ray, radio (Sunyaev--Zel'dovich observations), and optical (for gravitational lensing and dynamics). Our best model for El Gordo assumed initial total masses of $1.4 \times 10^{15}$~${\rm M}_{\odot}$ and $7.5 \times 10^{14}$~${\rm M}_{\odot}$ for the main and the infalling cluster respectively, an impact parameter of 300 kpc, and a relative initial infall velocity of 2250 km/sec when separated by the sum of the two virial radii. This model explains most of the observed features of El Gordo: the distinctive cometary feature with a  {\it twin-tailed} wake observed in the X-ray morphology, the locations of the two peaks of the dark matter components, and the position of the SZ peak. In this paper we use the total mass distribution from our best model to derive the surface mass distribution, and compare that to the surface mass distribution we derived from gravitational lensing.

\subsection{Multiply lensed galaxies}
%---------------------------
In this section we discuss the lensed systems used to constrain the mass model. 
The identification of multiply lensed galaxies systems in El Gordo is particularly challenging, particularly because no multiply lensed galaxies with spectroscopic redshifts have been confirmed so far, despite the recent attempts made with LDSS3 in Magellan by \cite{Cerny2018}.
However, with the increased depth of the new RELICS data, we can improve upon previous identifications. As a default, we adopt the original naming scheme of \cite{Zitrin2013} for the lensed system candidates, who identified the first families of strongly lensed galaxies, and performed the first strong lensing analysis of this cluster based on the three HST bands available at the time (the compilation of systems is detailed in appendix A). 
Notably, two systems (1 and 2 in Fig.~\ref{Fig_CC_Arcs_Chandra}) exhibit well resolved morphological features that, together with robust photometric redshifts, allow to unambiguously confirm these two families of images. Systems 3, 4 and 5 contain also morphological information and reliable photometric redshifts, which makes the identification of these systems equally robust. Systems 1 through 4 are similar to the systems defined in \cite{Zitrin2013}, and \cite{Cerny2018}. We note that systems 10 and 20 in \cite{Cerny2018} are part of our systems 1 and 2, where we identify different knots in the systems that are used as additional constraints. Our new system 5 was also independently identified by \cite{Cerny2018} as system 13 in their work. In the original scheme of \cite{Zitrin2013}, system 4 was composed of 3 tangential counterimages, and two possible radial images. A preliminary model clearly disfavors the radial image 4.2 and candidate image 4.3 \citep[in][]{Zitrin2013} as part of system 4. We note that \cite{Cerny2018} also rejected these two counterimages as part of system 4 (as do some updated models by Zitrin et al.; \emph{private communication}).
Instead, we suggest that the two alleged arclets 4.2 and 4.3 in \cite{Zitrin2013} are likely features in the galaxy cluster associated with the cooling of the plasma (see Section\ref{Sect_Filament}). 
Using the robust systems 1 through 5, we derive a first model based on these reliable systems. This model is later used to unveil new system families (listed in Table~\ref{tab_arcs}). We refer to this first model for the mass distribution as the {\it driver model}.\\

Table~\ref{tab_arcs} lists all the arclets, including also some candidates listed here for completeness, but not used in the lens reconstruction.  Also for completeness, we include system 8 as originally defined in  \cite{Zitrin2013}. The driver model disfavors this system, so we do not include it in our lens reconstruction. Also  \cite{Cerny2018} discarded this system. The systems in table~\ref{tab_arcs} are divided in 3 categories, A, B and C. Arclets in category A are robustly confirmed based on their color, morphology and photometric redshift. As mentioned above, we use these arclets to derive the driver model. Systems in category B are highly compatible with the driver model. In addition, color, and when available, morphology and photometric redshift is also consistent among the different members of the same family of images. Note that in some cases, the images are unresolved so they lack detailed morphological information. Also, gaps in the CCDs, or masked regions in some bands may result in color artifacts. Systems labeled A and B are used to derive an alternative model that we name the {\it full model}. Arclets in the category C are still consistent with the driver model, but lack of morphological information, a mismatch in the alignment of the predicted image (compared with the observed one), or tension between the predicted and observed magnification ratios reduces the reliability of the identification. Arclets marked with label C are not used in the mass reconstruction, but are still included in table~\ref{tab_arcs}. Future data will confirm or reject these system candidates. 

The systems in table~\ref{tab_arcs} that are new identifications are marked with bold face. Systems that were fully included in previous work are indicated in the {\it Comments} column. Our new system 6 has an estimated redshift (from the lens model) of $z \approx 4.3$, which is consistent with the photometric redshift. 
For system 7, we identify a new candidate for 7c that differs from the candidate in \cite{Zitrin2013}. 
System 10 is a new system with a photometric redshift of 5.1 (for 10a). The driver model is fully consistent with this system and redshift. 
System 11 is a new redefinition of system 5 in \cite{Zitrin2013}. The driver model suggests that the big arclet forming part of system 5 in \cite{Zitrin2013} consists of two images merging at the critical curve. The corresponding third image is identified with the tail of a bright galaxy, that may itself being lensed by another member galaxy (see Figure \ref{Fig_System11}). Based on the driver model, the redshift of this galaxy should be $z \approx 3.1$ while the photometric redshift for the arc is $z \approx 2.2$. When this system is included in the lens reconstruction (i.e in the full model), we adopt the photometric redshift for this system. 
System 12 is a redefinition of system 14 in \cite{Cerny2018}, based on the driver model and color+morphological information. Our 12a matches 14a in \cite{Cerny2018}, but we identify two different counterimages. The driver model predicts a redshift of $z \approx 3$, consistent with the photometric redshift of $z=3.4$ of 12a. 
System 13 has no photometric redshift. The driver model predicts a redshift $z=3$ for this system. 
System 14 has a photometric redshift $z=2.7$ (14a), but we adopt the redshift predicted by the driver model, i.e $z=4$, for this system. 
System 15 corresponds to system 8 in \cite{Cerny2018}, which was also independently identified. Both the photometric redshift ($z=2.7$), and the redshift predicted by the driver model ($z=2.65$) agree reasonably well.
System 17 is a newly identified system with two identifiable knots, and with a redshift consistent with the nearby system 3, so systems 3 and 17 may be forming a close pair of galaxies at $z \approx 5$. 
Systems 18 and 19 are all new candidates, but lack of morphological features do not allow us to confirm their association based on the morphology of the predicted images.

Finally, we do not consider system 5 in  \cite{Cerny2018}. Although the driver model is consistent with the positions of  system 5 in  \cite{Cerny2018}, a third image is clearly predicted, but not observed, casting doubt on the feasibility of this system. However, we should note that it is also possible that the driver model fails at correctly predicting the position of the third counterimage, or that this image could be hidden underneath one of the bright member galaxies, or that a small unobserved substructure close to the predicted position of this third image demagnifies this image. \\

We emphasize that, although no spectroscopic redshifts are yet available for this cluster, the photometric redshifts of the systems used to derive the driver model show a high degree of consistency and relatively small scatter. The driver model can then geometrically predict the redshifts of the remaining system candidates \citep[see for instance][for an application of this method]{Chan2020}. Future spectroscopic observations will confirm some of these systems, and/or discard others, but will allow also to test the blind predictions made by the driver model, hence validating the process of estimating redshifts in a geometric way.
%--------------------------------------------------------------------
\section{Formalism}\label{sec_math}
%%%%%%%%%%%%%%%%%%%%%%%%%%%%%%%%%%%%%%%%%%%%%%%%
The mass reconstruction is based on our method WSLAP+. The reader can find the details of the method in our previous papers
\citep{Diego2005,Diego2007,Sendra2014,Diego2016}. Here we give a brief summary of the most essential elements. \\
The lens equation is defined as follows, 
\begin{equation} 
\beta = \theta - \alpha(\theta,\Sigma), 
\label{eq_lens} 
\end{equation} 
where $\theta$ is the observed position of the source, $\alpha$ is the deflection angle, $\Sigma(\theta)$ is the surface mass-density of the
cluster at the position $\theta$, and $\beta$ is the position of the background source. Both the strong lensing and weak lensing
observables can be expressed in terms of derivatives of the lensing potential:\footnote{Note however, that through observations one measures the reduced shear, $\gamma_r=\gamma/(1-\kappa)$ where $\kappa$ is the convergence.}
\begin{equation}
%\label{2-dim_potential} 
\psi(\theta) = \frac{4 G D_{l}D_{ls}}{c^2 D_{s}} \int d^2\theta' \Sigma(\theta')ln(|\theta - \theta'|), 
%\label{eq_psi} 
\end{equation}
where $D_l$, $D_s$, and $D_{ls}$ are the angular diameter distances to the lens, to the source and from the lens to the source, respectively. The unknowns of the lensing problem are, in general, the surface mass density, and the positions of the background sources in the source plane. 
The surface mass density is described by the combination of two components; 
i) a soft (or diffuse) component (usually parameterized as superposition of Gaussians); and 
ii) a compact component that accounts for the mass associated with the individual halos (galaxies) in the cluster. \\
For the diffuse component, different bases can be used but we find that Gaussian functions provide a good compromise between the desired compactness and smoothness  of the basis function. A Gaussian basis offers several advantages, including a fast analytical computation of the integrated mass for a given radius, a smooth and nearly constant amplitude between overlapping Gaussians (with equal amplitudes) located at the right distances, and a orthogonality between relatively distant Gaussians that help reduce unwanted correlations. 
%However, we should note that the choice of basis is not fee of arbitrariness and that other basis functions can be used instead of Gaussians \citep[see][for a discussion of alternative basis functions including polynomial, isothermal models or power laws]{Diego2007}. 
For the compact component, we adopt directly the light distribution in the IR band (F160W) around the brightest member galaxies in the cluster. 
For each galaxy, we assign a mass proportional to its surface brightness. This mass is later re-adjusted as part of the optimization process. 
%The compact component is usually divided in independent layers, each one containing one or several cluster members. The separation into different layers allows us to constrain the mass associated to special halos (such as the giant elliptical galaxies) independently from more ordinary galaxies. This is useful in the case where the light-to-mass ratio may be different, like for instance in the BCG. \\

%%----------------------------------------------------------------- 
%   \begin{figure}
%   \centering
%   \plotone{Figs/ElGordo_MassContour_vs_Optical.pdf}
%      \caption{Contours of the mass distribution for the full model compared with the optical image. The circles mark the position of the BCG in the SE and the center of the NW group.   
%              }
%         \label{Fig_Mass_Optical}
%   \end{figure}
%%-----------------------------------------------------------------

As shown by \cite{Diego2005,Diego2007}, the strong and weak lensing problem can be expressed as a system of linear
equations that can be represented in a compact form, 
\begin{equation}
\vectTheta = \matrGamma \vectX, 
\label{eq_lens_system} 
\end{equation} 
where the measured strong lensing observables (and weak lensing if available) are contained in the array $\vectTheta$ of dimension $N_{\Theta }=2N_{\rm sl}$, the unknown surface mass density and source positions are in the array $\vectX$ of dimension 
\begin{equation}
N_{\rm X}=N_{\rm c} + N_{\rm g} + 2N_{\rm s}
\label{eq_Nx}
\end{equation}
and the matrix $\matrGamma$ is known (for a given grid configuration and fiducial galaxy deflection field) and has dimension $N_{\Theta }\times N_{\rm X}$.  $N_{\rm sl}$ is the number of strong lensing observables (each one contributing with two constraints, $x$, and $y$), and $N_{\rm c}$ is the number of grid points (or cells) that we use to divide the field of view. Each grid point contains a Gaussian function. The width of the Gaussians are chosen in such a way that two neighbouring grid points with the same amplitude produce a horizontal plateau in between the two  overlapping Gaussians. In this work, we consider only regular grid configurations. Irregular grids are useful when there is a clear peak in the mass distribution, for instance when the cluster has a well defined centre or a single BCG. 
$N_{\rm g}$ (in Eq.~\ref{eq_Nx}) is the number of deflection fields (from cluster members) that we consider. 
$N_{\rm g}$ can be seen as a number of mass layers, each one containing one or several galaxies at the distance of the cluster.  
In this work we set $N_{\rm g}$ equal to 1, i.e, all galaxies are forced to have the same mass-to-light ratio.
%In the case where  $N_{\rm g}=1$, all the individual galaxies in the lens model are assumed to follow the same light-to-mass ratio and are re-scaled by the same parameter (that is, they are all in the same layer). In a second scenario we assume $N_{\rm g}=2$ where all galaxies are in the same layer except the BCG that is in the southern part of the cluster (near the giant arc). The reason for this configuration is that the northern BCG seems to be poorly constrained by the lensing data so by adopting $N_{\rm g}=2$ we can explore the case where the mass-to-light ratio of the northern BCG is fixed together with the cluster members' mass-to-light ratio and we let the southern BCG be constrained by the lensing data. In the case where  $N_{\rm g}=3$, each BCG is allowed to have its own mass-to-light ratio and the remaining galaxies are placed in the third layer (and hence forced to have the same mas-to-light ratio). $N_{\rm g}=3$ results in models where the northern BCG contains a significantly larger mass (a factor $\approx 2$ larger) and predicts new arcs that are not observed in the data making this model less favored by the data. The particular configuration of the galaxies in our lens model is shown in figure \ref{fig_contour}. 

%%----------------------------------------------------------------- 
   \begin{figure*}
   \centering
   \plottwo{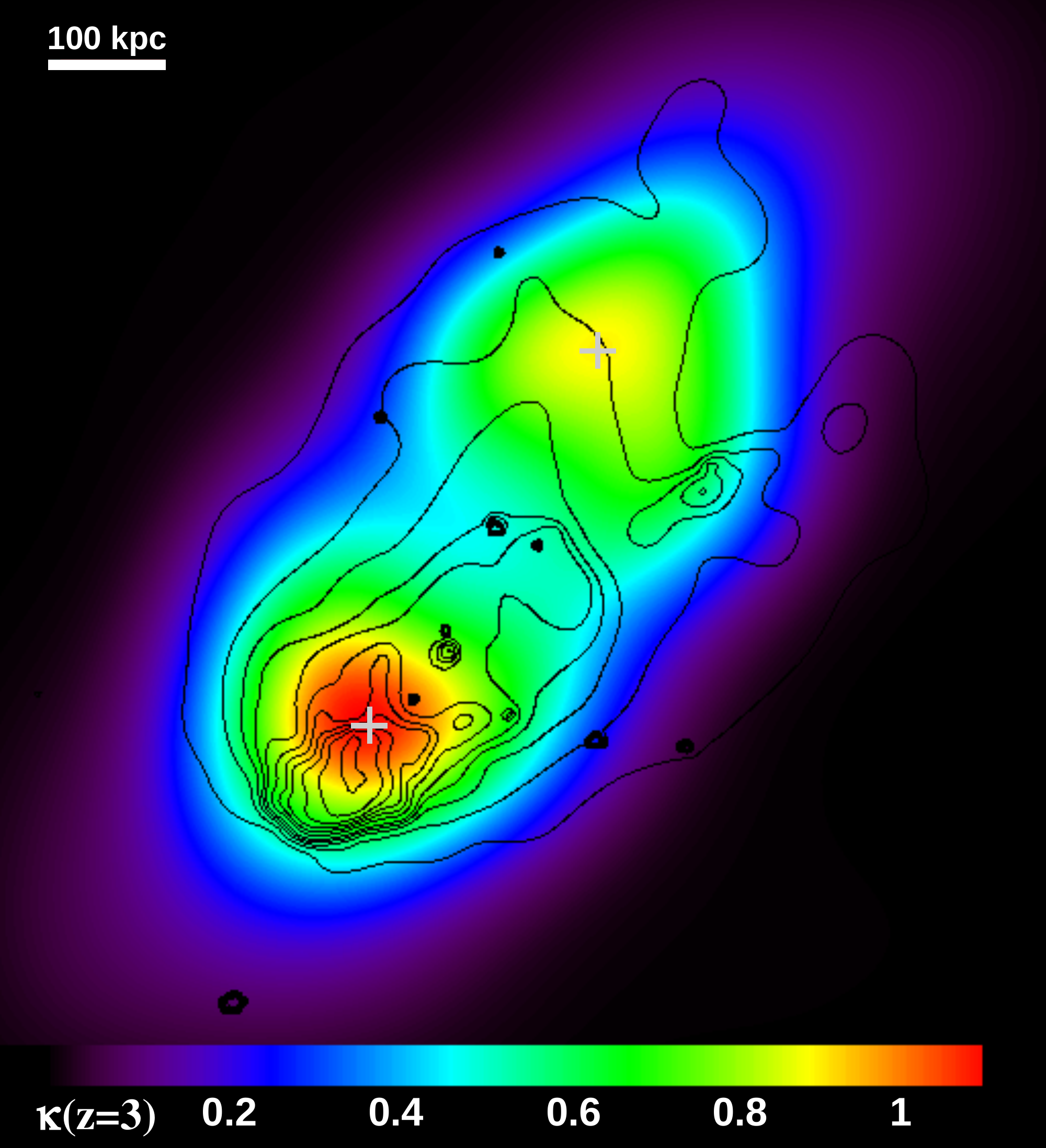}{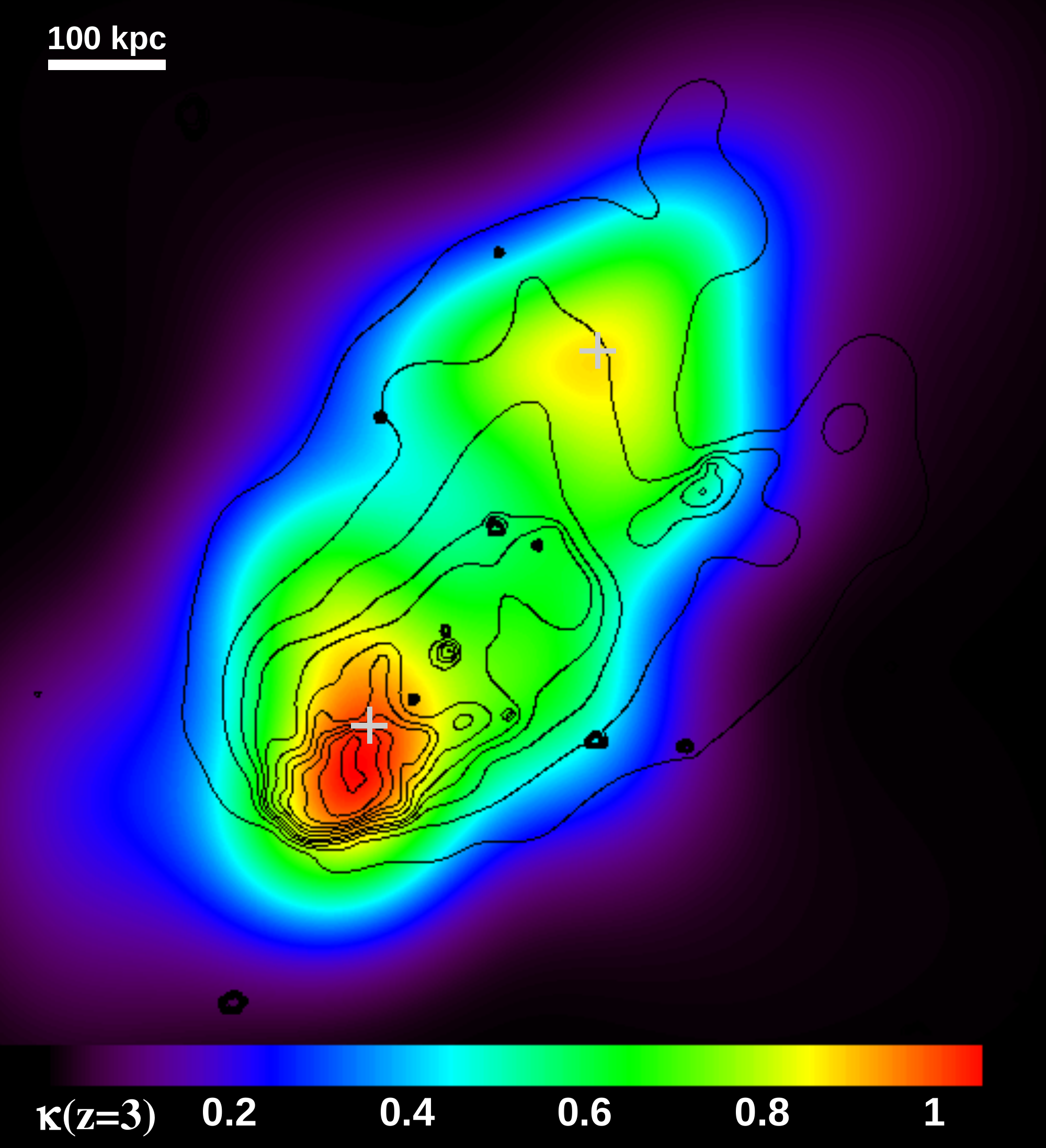}
      \caption{Left panel. Convergence (color scale at z=3) vs X-ray contours (contours) for the driver model. The converge has the contribution from the member galaxies removed. 
               The crosses mark the position of the BCG (or SE group) and NW group. Right panel. Similar as the left panel, but for the full model. 
               Note the good correlation between the X-ray and SE mass peak for the full model.  
              }
         \label{Fig_Mass_Xray}
   \end{figure*}
%%-----------------------------------------------------------------

Finally, $N_{\rm s}$ in Eq.~\ref{eq_Nx} is the number of background sources (each contributes with two unknowns, $\beta_x$, and $\beta_y$), which in our particular case ranges from $N_{\rm s}=5$ when only the subset of reliable systems are used (driver model in section~\ref{sect_data}) to $N_{\rm s}=16$, when all systems labeled A or B in Table~\ref{tab_arcs} are used in the reconstruction (full model). The solution, $X$, of the system of equations \ref{eq_lens_system} is found after minimizing a quadratic function of $X$ \citep[derived from the system of equations \ref{eq_lens_system} as described in ][]{Diego2005}. The minimization of the quadratic function is done with the constraint that the solution, $\vectX$, must be positive. Since the vector $\vectX$ contains the grid masses, the re-normalisation factors for the galaxy deflection field and the background source positions, and all these quantities are always positive (the zero of the source positions is defined in the bottom left corner of the field of view).  Imposing  $\vectX>0$ helps constrain the space of meaningful solutions, and to regularise the solution, as it avoids unwanted large negative and positive contiguous fluctuations. The quadratic algorithm convergence is fast (a few minutes on a standard laptop), allowing for multiple solutions to be explored in a relatively short time. Different solutions can be obtained after modifying the starting point in the optimization and/or the redshifts of the systems without spectroscopic redshift. A detailed discussion of the quadratic algorithm can be found in \cite{Diego2005}. For a discussion of its convergence and performance (based on simulated data), see \cite{Sendra2014}.
In order to study the dependency of the lens model with the uncertainty in the photometric redshifts and the initial condition, we derive also 100 lens models with the settings of the driver model (i.e using A systems only). For each one of the models we vary the initial condition in the optimization process and the redshift of all five background sources. For the redshift we adopt an error of 0.5 for all 5 systems, which is consistent with the dispersion in photometric redshifts for each system as shown in table~\ref{tab_arcs}.

\section{Results}\label{sect_result}
%%%%%%%%%%%%%%%%%%%%%%%%%%%%%%%%%%%%%
Thanks to the new RELICS data, we can revise the multiple images identification in this cluster and assign them a rank ranging from A (most reliable) to C (least reliable). Based on the set of images ranked A (see Table~\ref{tab_arcs}) we derive the driver model, which is later used to uncover new multiply images, or to reveal issues with previous identifications. Even though the driver model is based on a relatively small subset of only 5 families of images, the relatively uniform spatial distribution of these 5 families allows us to derive a relatively reliable lens model. The driver model disfavors the radial counterimages candidates 4.2 and 4.3 in \cite{Zitrin2013} \citep[these images were discarded also by][]{Cerny2018}, and instead we suggest these may be signatures of cooling flows or jets near the BCG (see Section \ref{Sect_Filament} for discussion). System 8 in \cite{Zitrin2013} shows a relative good consistency with the driver model in terms of predicted vs observed positions, but the morphology of the observed images does not match well with the predicted morphology, so we do not use this system in any of our lensing reconstructions as well (this system is still included in Table~\ref{tab_arcs} for completeness). 
Some of the counterimages postulated in earlier work as candidates (for instance 7c and 9c) are in general consistent in terms of position, but their morphology is not well reproduced by the lens model. We also unveil new image candidates, some of them independently identified in \cite{Cerny2018}. 

In addition to these, we identify additional new families as described in the appendix. System 15 in \cite{Cerny2018} is consistent with the driver model, but a third image is clearly predicted by the driver model and not observed. Consequently, we do not use this system in our reconstruction, although we should note that the predicted position for the third image is only a few arcseconds from the BCG. Hence, it is possible that the driver model is not accurate enough around the BCG, and that the third image lurks behind the bright BCG, and with a smaller magnification than the one predicted by the driver model. A smaller magnification is possible if the BCG has a larger mass-to-light ratio in the central region, for instance through a central spike in the mass distribution, or if a supermassive black hole is at, or near, the centre. 

Based on the driver model, we expand the number of reliable systems and estimate their redshifts based on the available photometric redshift information and/or the redshift predicted by the driver model. Using the expanded set of systems (ranked A and B in Table~\ref{tab_arcs}), we derive a new model, which we refer to as the {\it full} model. The mass distributions of the two models are compared in Figure~\ref{Fig_Mass_Xray}. For these plots, we have subtracted the contribution from the member galaxies to better show the diffuse component. The two models look similar to first order, but some differences can be appreciated specially around the BCG, where the full model places the peak of the diffuse component at several arcseconds from the BCG. In particular, the peak of the soft component correlates very well with the peak in the X-ray emission. Similar correlations between the diffuse component obtained by our method and the observed X-ray emission were found in earlier work, where we discussed the possibility that the lensing data is sensitive also to the mass of the hot gas. This possibility is discussed in more detail in section \ref{Sect_Filament}. Note that, in massive clusters, the hot plasma is expected to contribute on average to $\approx 10\%$ of the total mass. This fraction can increase locally, specially in those central regions where the cooling is more efficient due to the square dependency of the X-ray emission with the electron density. 

\begin{figure}
   \centering
%%   \plotone{Figs/Mass_profile_ElGordo_V2.pdf}
   \plotone{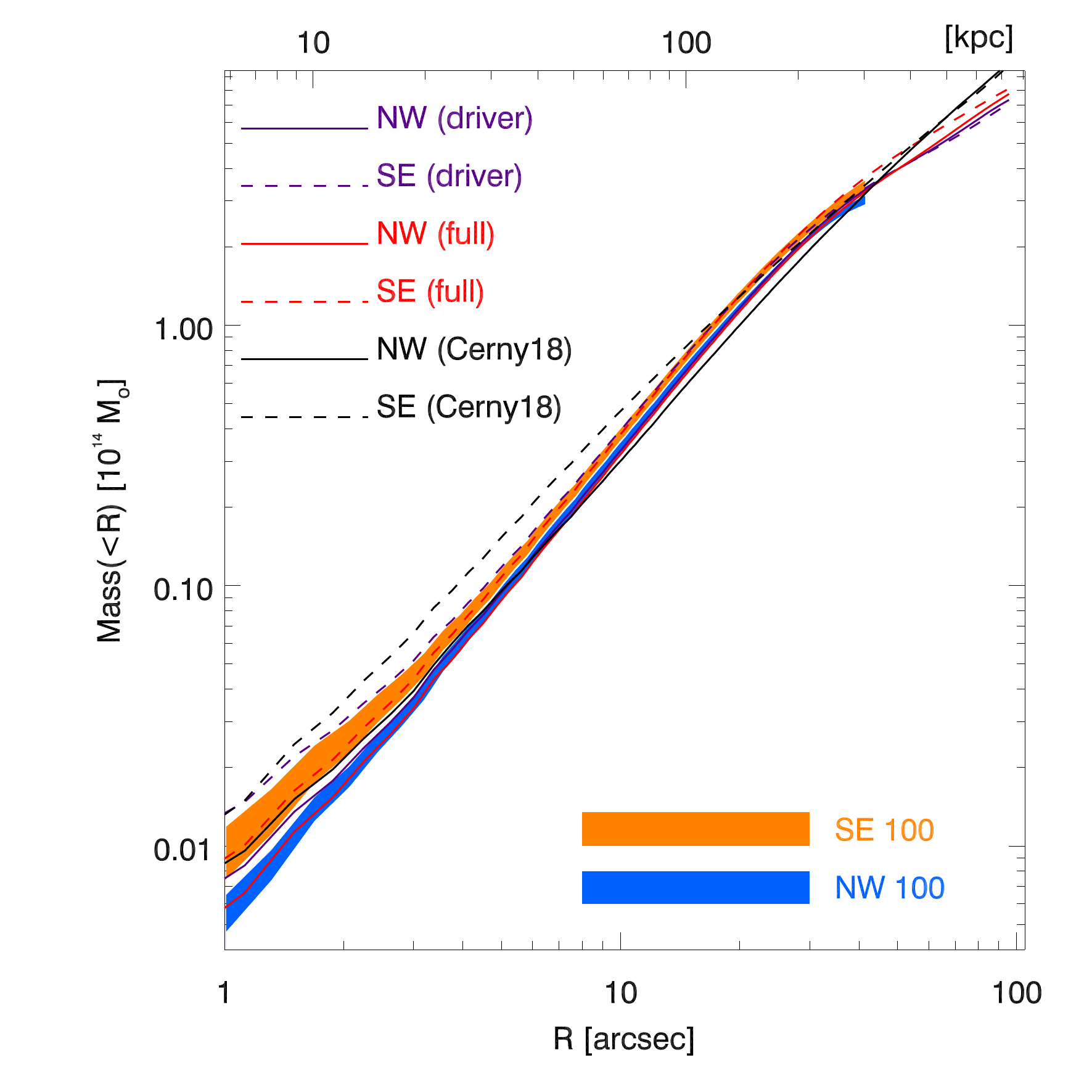}
      \caption{Integrated mass as a function of aperture radius. The purple lines correspond to the driver model, the red lines correspond to the 
               model derived with all arcs. For comparison we show also the corresponding integrated mass for the mass model from \cite{Cerny2018} as orange lines 
               In all cases, the solid lines are for profiles centered in the NW group while the dashed lines are for profiles centered in the SE BCG. The colored regions show the 1-$\sigma$ interval derived from the 100 realizations, where both the initial conditions and photo-z are varied.
              }
         \label{Fig_Profiles}
   \end{figure}

In terms of integrated mass, both models look also similar, but with the full model having slightly more mass, specially on the SE group where the X-ray emission is also more intense. A quantitative comparison of the integrated mass as a function of radius for each subgroup and the two models is presented in Figure~\ref{Fig_Profiles}. Here, we include also the dispersion of the 100 driver models produced after varying the initial condition and photometric redshifts (colored bands).
The mass increase in the full model around the SE group is due mostly to the smaller photometric redshift of system 11 used to derive the full model compared with the larger redshift predicted by the driver model. Figure~\ref{Fig_System11} shows the predicted images for 11a and 11b based on 11c for the driver model and the full model. The driver model does a good job at predicting the arc position and morphology for a source redshift of $\approx 3$. In the full model, the system is assumed to be at the photometric redshift of $z=2.2$ instead, which results in a mass increase in the SE group needed to compensate for the smaller redshift of the background source.

%%----------------------------------------------------------------- 
%%-----------------------------------------------------------------

\begin{figure}
   \centering
   \plotone{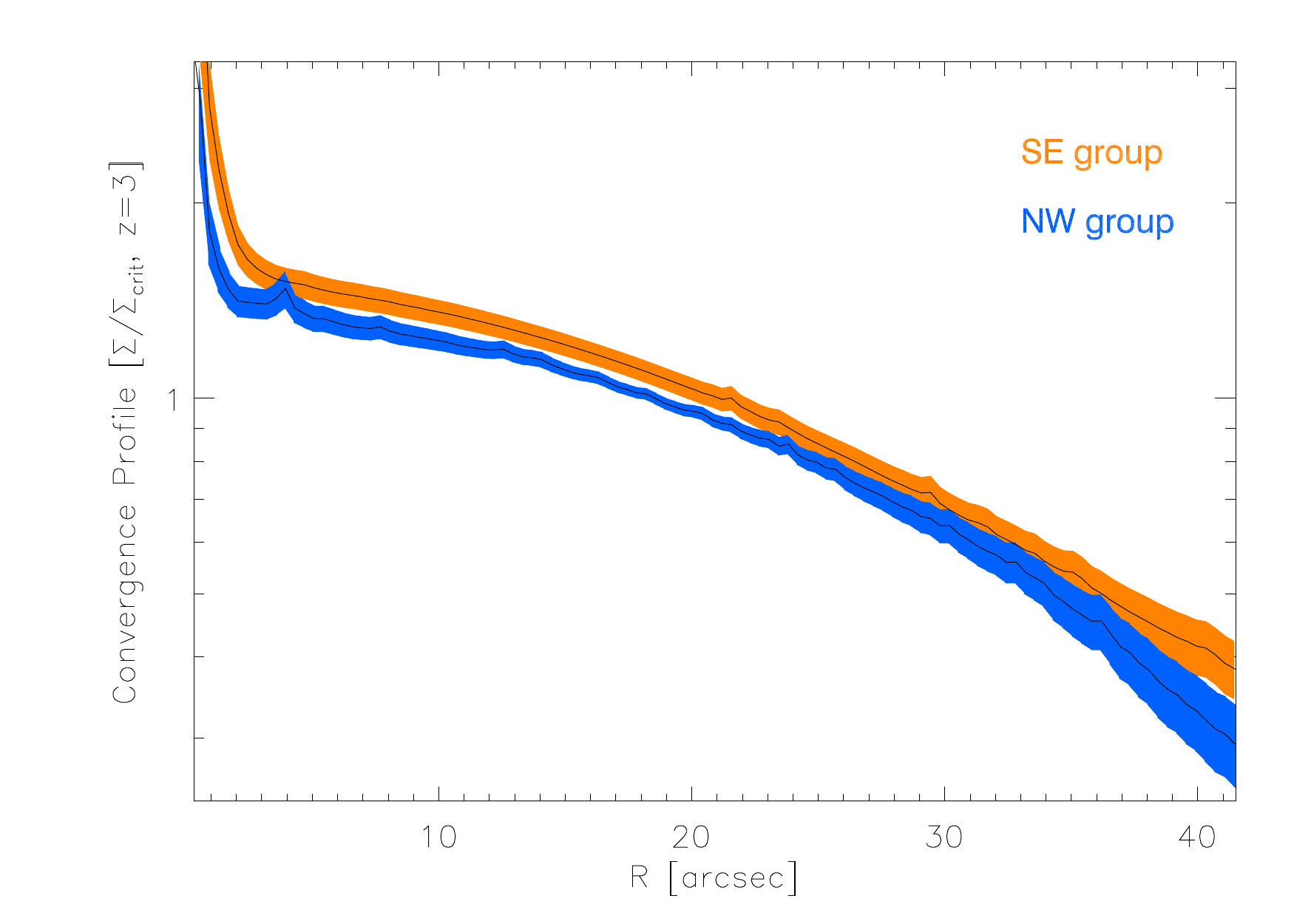}
      \caption{Average surface mass density profiles of the 100 realizations where both the initial condtitions and photo-z are varied. The colored bands show the 1-$\sigma$ interval.
              }
         \label{Fig_Profiles_2D}
   \end{figure}

\begin{figure}
   \centering
   \plotone{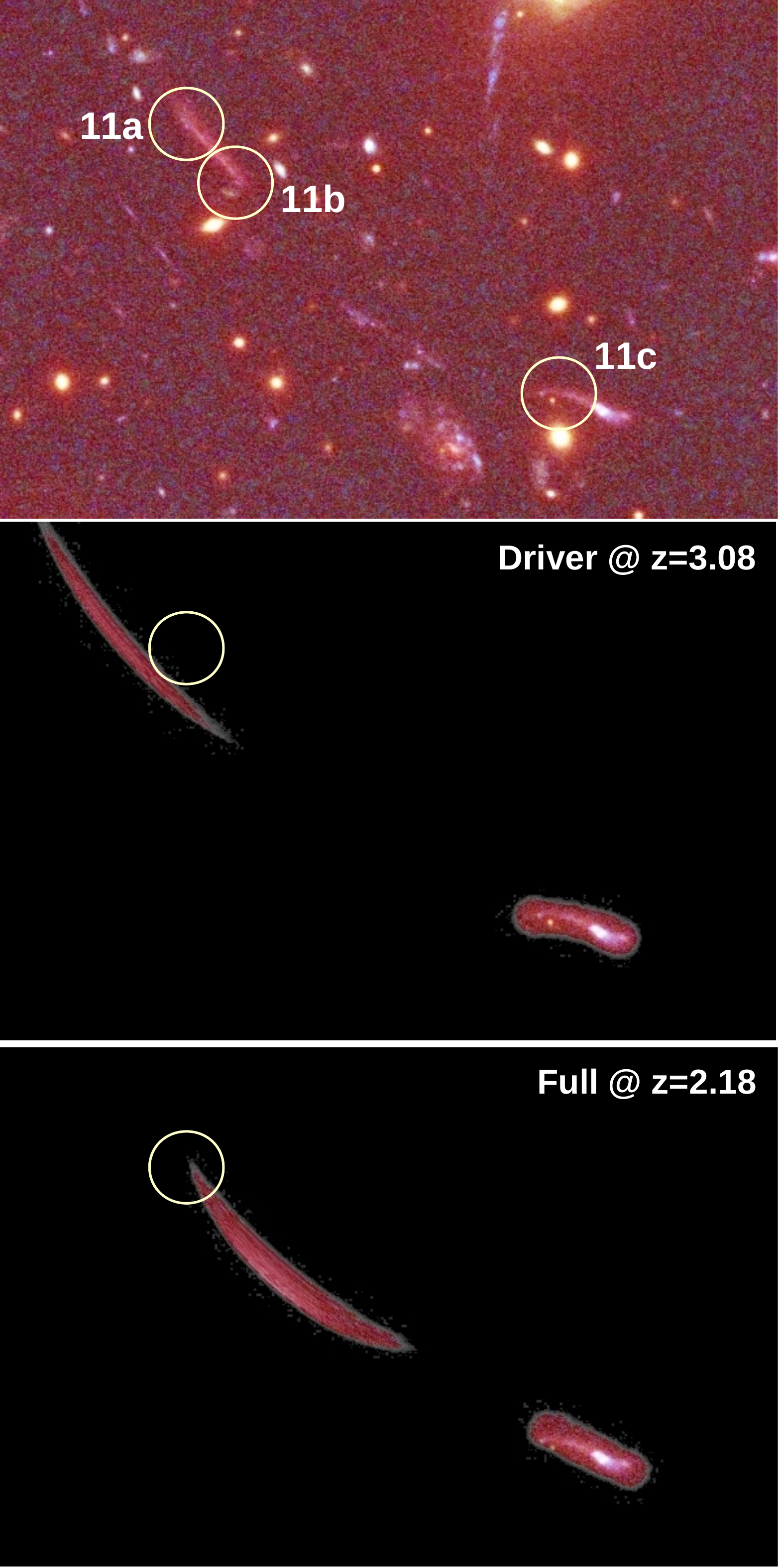}
      \caption{The top panel shows the proposed redefinition of the original system 5 in \cite{Zitrin2013} as the new system 11. The middle panel shows the predicted merging arc by the driver model assuming the background galaxy is at z=3.08. The bottom panel shows the corresponding prediction made by the full model when the source is forced to be at the photometric redshift. The white circle in the middle and bottom panel marks the position of 11a. 
              }
         \label{Fig_System11}
   \end{figure}

In Figure~\ref{Fig_Profiles_2D} we show the dispersion of the radial profiles for the 100 driver models derived after varying the initial condition and photometric redshifts. The profiles are derived after centering around the NW and SE group, and requiring that the mass of one halo is not included in the other halo. We impose this condition by masking one half of the field of view (through the diagonal) and computing the average profile in the unmasked region. Clearly, the SE group is more massive than the NW group, in agreement with the results of \citep{Zitrin2013,Cerny2018}.

%%----------------------------------------------------------------- 
%%-----------------------------------------------------------------

\subsection{Comparison with earlier results}
%%%%%%%%%%%%%%%%%%%%%%%%%%%%%%%%%%%%%%%%%%%%%%%%
In this section we compare our models with previous results derived from the same RELICS data and presented in \cite{Cerny2018}. We should note that the constraints used in this work and in \cite{Cerny2018} are not exactly the same, so some of the differences can be attributed to this fact. Our lensed candidates were derived independently from \cite{Cerny2018}, although in some cases, our system candidates coincide, but not in all cases. 
System 3, with photometric redshift $z=7.42$ in \cite{Cerny2018}, is claimed to be a newly identified system. However, the positions of 3.1, 3.2 and 3.3 are very similar (within a fraction of an arcsec) to the positions of system 3 in \cite{Zitrin2013}, but in this case, with a different photometric redshift of 4.16. For the position and redshift of system 3, we adopt the values of \cite{Zitrin2013} based on the color of the images and positions relative to the critical curve of preliminary models. We have not used system 8 from \cite{Cerny2018}. System 8 appears as a likely multiple lensed arc with two pairs of images very close to each other and possibly merging around a critical curve. A third predicted image can not be found near the predicted position suggesting that this is not a real system, or that the assumed redshift is significantly different than the real one. Alternatively, the predicted image may be buried behind a member galaxy. System 10 in \cite{Cerny2018} is included as an extra knot in our system 1. Similarly, we have included the positions of system 20 of \cite{Cerny2018} as additional knots of system 2. 
For the driver model, all systems are included also in \cite{Cerny2018} (although, see the difference in redshift for system 3). For the alternative full model, we include all systems used in \cite{Cerny2018}, except system 8 as mentioned above. In the full model, we also include systems not included in \cite{Cerny2018}. These are the new system candidates 6, 10, 13, 14, 16, and 17 and the redefined system candidates 11, 12 and 14 listed in the table in appendix. The system 14 in \cite{Cerny2018} was not included in their model. Here we use a redefined version of this system as our new system 12.

%%----------------------------------------------------------------- 
%%-----------------------------------------------------------------

\begin{figure}
   \centering
   \plotone{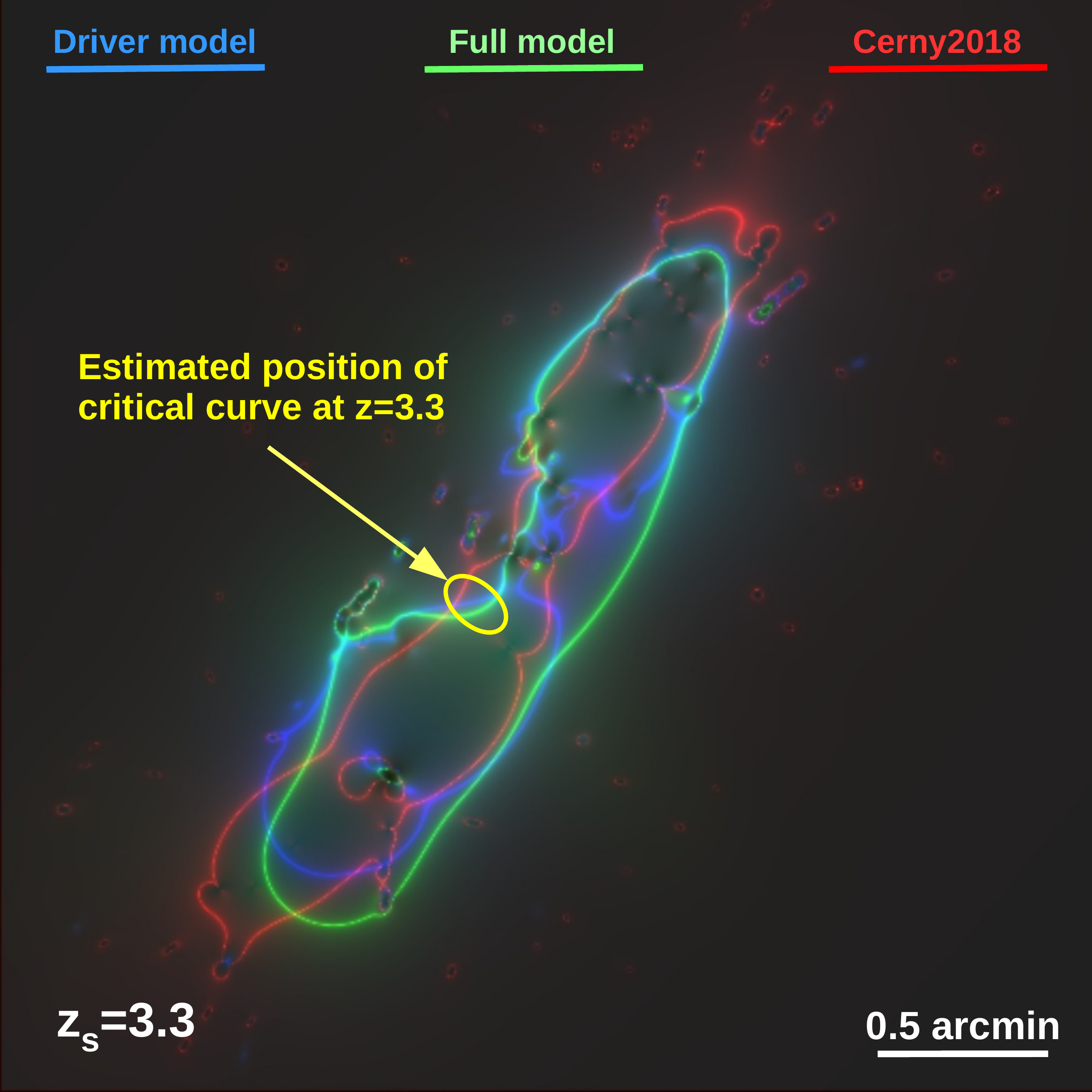}
      \caption{Comparison of the critical curves between our models and the model in \cite{Cerny2018}. The yellow ellipse marks the observed position of the critical curve from system 2 (at $z \approx 3.3$). 
              }
         \label{Fig_CCcomparison}
   \end{figure}

The critical curves of our two models and the model in \cite{Cerny2018} are compared in Figure~\ref{Fig_CCcomparison}. The position of the critical curves is consistent between both models, although our model predicts slightly wider critical curves, suggesting a rounder distribution for the projected surface mass density in our model. In contrast, \cite{Cerny2018} predicts a more elongated distribution of matter,  with the mass being more concentrated around the line intersecting the two clumps. The models show a better agreement (in terms of positions of the critical curve) around the position of the constraints. The figure shows the estimated observed position of the critical curve based on symmetry arguments for the giant arc of system 2 (at $z\approx 3.3$). All models agree relatively well with this position by placing the critical curve (at the redshift of system 2) very close, or intersecting, the estimated position of the critical curve. In the South-East part of the lens, differences between models are larger, reflecting the relative smaller density of constraints in this part of the lens (see Figure~\ref{Fig_CC_Arcs_Chandra}), but possibly also the fact that parametric methods assume explicit mass profiles that can extend the mass distribution beyond the range of distances covered by the lensing constraints. A more quantitative comparison of the magnification between the different models can be made by comparing the curves, $A(>\mu)$, of the area above a given magnification. These curves are computed by integrating the differential area curves, i.e  $A(>\mu) = \int_{mu}^{\mu_{max}}d\mu dA/d\mu$ where $\mu_{max}$ is the maximum magnification considered (220 in this case) and $dA/d\mu$ is the area in the lens plane with magnification $\mu$ and in the interval $d\mu$, divided by the magnification $\mu$ (i.e, the corresponding area in the source plane). The curves  $A(>\mu)$ follow the usual $A_o\mu^{-2}$ above magnification $\mu\approx 10$. The values of the normalization for the different models and at $z_s=3.3$ are (in arcmin$^2$): $A_o=4.5$ (Cerny18 model), $A_o=10$ (driver model) and $A_o=8.5$ (full model).  $A(>\mu)$ can be interpreted as the probability of a galaxy being lensed by a factor larger than $\mu$. 
At high magnifications, the driver and full models predict about twice the probability compared with the model in \cite{Cerny2018}. This difference is mostly due to the shallower profiles in the driver and full models around the position of the critical curves. The values of $A_o$ put El Gordo at a level comparable to other powerful lenses, like the Hubble Frontier Fields clusters, in terms of lensing efficiency \citep[see for instance][]{Vega-Ferrero2019}. This means that future observations, like the planned ones with JWST, promise to reveal many additional high-redshift lensed galaxies.

Due to the relatively large separation between the two groups, the critical curve on the Central-East side of the cluster is relatively unperturbed by cluster members (this can be appreciated in Figure~\ref{Fig_CC_Arcs_Chandra} where the critical curves are very smooth in this part of the cluster). At even higher redshift, the critical curves move outwards where the distortion by cluster members is expected to be even smaller. This has important implications for the probability of observing caustic crossing events of distant stars, for instance Pop III stars at $z>7$ as suggested by \cite{Windhorst2018}. Pristine critical curves (that is, critical curves which are not perturbed by microlensing from stars or remnants in the cluster members or in the intracluster medium) can produce lensed images in their vicinity with magnifications factors of order $10^6$ when the background source has the size of a Pop III star. In contrast, critical curves that are close to the cluster centre (for instance, for background objects at relatively low redshifts of $z\approx 2$ or less) are normally perturbed by such microlenses resulting in maximum magnification factors of order $10^4$ for background sources with sizes comparable to giant stars \citep[see for instance][for details]{Diego2019}. The relatively unperturbed sections of the critical curves in El Gordo are hence good target regions for JWST to search for highly magnified distant stars.

In terms of total mass, the agreement between the models is made more evident when looking at the integrated mass as a function of aperture radius. In order to better account for the asymmetric nature of the cluster, we set the centre of the aperture at the position of the two main galaxies (or BCGs). For each centre, we compute the projected mass within a given aperture as a function of the aperture radius. The resulting profiles are shown in Figure~\ref{Fig_Profiles}. All models agree well specially between $\approx 100$--$300$ kpc, which is the range where lensing constraints are more abundant. At small radii ($r<100$ kpc), the model in \cite{Cerny2018} predicts slightly more mass than our free-form models, specially in the SE clump. At radii larger than $\approx 400$ kpc our free-form models fall below the prediction of the model of \cite{Cerny2018}. This is an expected behaviour, since the free-form models have systematically low masses in areas that extend beyond the realm of the lensing constraints. This is simply a memory effect of the algorithm that does not constrain distant regions in the field of view, leaving their masses close to their initial value before the minimization (these masses are originally assigned small random values). 

\section{Discussion}\label{sect_disc}
%%%%%%%%%%%%%%%%%%%%%%%%%%%%%%%%%%%%%%%%%%%%%%%%
The results from the previous section suggest that the mass in El Gordo is relatively well constrained in the inner 500 kpc region. Within this range, \cite{Cerny2018} finds that the masses within the 500 kpc radius for each clump have a mass ratio of 1.19 (for SE/NW). Compared with our results, we find that within the same radius, the SE/NW ratio is 0.98 for the driver model and 1.11 for the model with all systems. At 100 kpc, this ratio grows to 1.17 and 1.18 for the driver model and full model, respectively. This should be compared with the dynamical masses inferred in \cite{Menanteau2012}, where for the SE/NW ratio (within the virial radius) they find a value of $0.6\pm0.4$, and hence consistent with a ratio of $\sim 1$ at $1\sigma$ with their measurement. The weak lensing analysis in \cite{Jee2014} finds a more discrepant ratio of the SE/NW groups (in the virial masses) of $0.56 \pm 0.17$ (statistical), in contrast with our results. This discrepancy may be due to systematic effects in either analysis, but it is also possible that the NW group becomes more massive than the SE group beyond the 500 kpc radius studied in the strong lensing analysis. 

One of the more puzzling aspects of El Gordo cluster is the position of the X-ray emission in relation to the peak in the mass distribution. \cite{Botteon2016} study this cluster with X-rays and infers a very high velocity for the shock (with a Mach number of 3 or above), which is spatially coincident with one of the radio relics. 
\cite{Ng2015} combines different observations from El Gordo cluster to constrain the dynamical state of the cluster. Based on the separation of the two subgroups, the morphology of the radio relics and their polarization angle, they conclude that the cluster is most likely in a return phase. This naturally explains the relative position of the X-ray peak and the main BCG, which seems to be lagging behind the X-ray peak. \cite{Hallman2004} studies the low-redshift cluster A168, which resembles El Gordo. Like in El Gordo, in A168 the peak of the X-ray emission seems to have moved ahead of the dominant galaxies in the cluster. They conclude that the ``subcluster gas slingshots past the dark matter center, becomes unbound from the subcluster and expands adiabatically''. This type of adiabatic expansion has been observed in N-body hydrodynamical simulations \citep{Mathis2005}, and it can result in a substantial cooling of the gas as it leaves the potential well. For a cluster in a return phase after a head-on collision, the gas leaves the potential well twice; a first time as it drags behind the peak of the mass distribution due to ram pressure, and a second time when the peak of the mass distribution falls back towards the centre of mass sprinting through the gas. 

The results from our driver model seems to agree better with the \cite{Ng2015} interpretation (returning phase), since the X-ray peak is ahead of the mass peak. On the contrary, the full model seems to agree  better with the interpretation of \cite{Molnar2015} and \cite{Zhang2015}, since the mass peak coincides with the X-ray peak, and not with the BCG. This would suggest that the BCG was perturbed out from the potential well of the infalling cluster, which may be explained by the merging.  Based on results from full N-body.hydro simulations \citep{Molnar2015,Zhang2015}, the infall velocity for El Gordo is not as large as for the Bullet cluster. If the velocity is not large enough, the ram pressure does not produce noticeable offsets between the mass and gas centers of the infalling cluster. Future spectroscopic confirmation of lensed systems, plus the addition of new systems from deeper JWST observations, will allow to resolve this question, by unambiguously determining the peak position of the mass in the SE group. A confirmation of the candidate systems (and their redshift) used in the full model would favour the scenario proposed by \citep{Molnar2015,Zhang2015}.

%I do not think anyone tried to simulate what happens with the BCG of an infalling cluster. It would be interesting.

%%----------------------------------------------------------------- 
%%-----------------------------------------------------------------

\subsection{Total mass}
\begin{figure}
   \centering
   \plotone{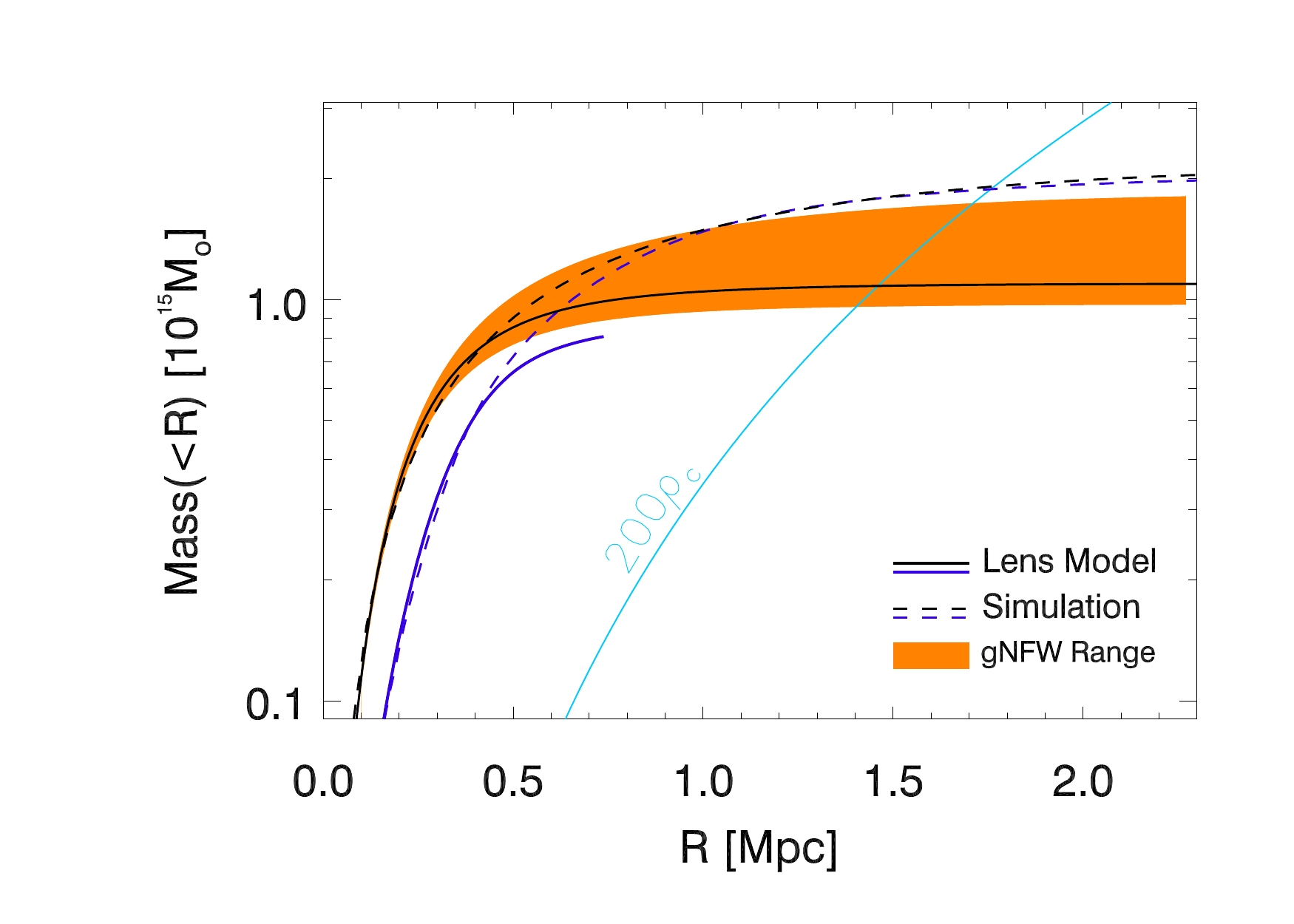}
      \caption{The thick black solid line shows the integrated mass of a double gNFW model fitting the driver lens model as a function of aperture radius (i.e mas projected in a cylinder with radius R). In the double fit, we place a gNFW in the centre of each subhalo. The solid blue line is the integrated mass of the driver model when the centre is in between the two clumps. The black-dashed line is the corresponding profile of a double gNFW profile fitting the simulated "El Gordo" cluster in \cite{Molnar2015} (i.e, like in the solid black curve), while the blue dashed line is the integrated mass as a function of aperture of the simulated cluster when the center is chosen in the middle of the two subgroups (i.e like in the solid blue curve). The light blue curve is the mass enclosed in a sphere with radius R and density 200 times the critical density (at the redshift of the cluster). The shaded orange region marks the range of double gNFW models that are consistent with the lens models. 
              }
         \label{Fig_TotalMass}
   \end{figure}

%-------------------------------
In this section we estimate the total mass of the lens model within $r_{200c}$ by simultaneously fitting two gNFW profiles to the lens model. We place one gNFW at the centre of each subclump and fit simultaneously the two profiles of the lens model. In this case, as opposed to what was shown in section \ref{sect_result} and Figures  \ref{Fig_Profiles}, \ref{Fig_Profiles_2D}, we do not mask the other subgroup when computing the profiles of each subgroup, since the simultaneous fit considers already the superposition of the two profiles in the intermediate region (and the other regions in the field of view).  
The gNFW is a generalization of the NFW profile and is given by;
\begin{equation}
\rho(r) = \frac{\rho_o}{(r/r_s)^{\gamma}[(1+r/r_s)^{\alpha}]^{(\beta-\gamma)/\alpha}}
\end{equation}
where $\rho_o$ is the normalization and $\gamma$, $\alpha$ and $\beta$ are the inner, middle, and outer slopes of the profile. In a standard NFW profile  $\gamma=1$, $\alpha=2$, and $\beta=3$. Since the lens model can only constrain the inner $\approx 300$ kpc, the slopes $\alpha$ and $\beta$ are unconstrained. Hence, we fit only for the normalization, $\rho_o$, inner slope, $\gamma$, and scale radius $r_s$. We fix $\alpha$ to the value for the standard NFW (i.e $\alpha=2$). For $\beta$ we consider two values, since the value of $\beta$ largely determines the integrated mass up to the virial radius. In particular, we consider the two cases of $\beta=3$ (NFW-like case) and  $\beta=2$ (isothermal-like case) to explore the uncertainty due to the extrapolation of the double gNFW profile up to the virial radius. While the case $\beta=3$ is consistently reproduced by N-body simulations, shallower values are expected at large radii when one considers contributions form the 2-halo term. We find that shallow values for the central slope of $\gamma=0.2$, together with scale radius in the range 285--380 kpc are needed in order to reproduce the mass distribution of the lens model. The small scale radii correspond to the isothermal-like profiles, $\beta=2$, while larger scale radii are needed when adopting the NFW-like profiles, $\beta=3$. By adding the two gNFW profiles and extrapolating to larger radii it is in principle possible to estimate the mass at the virial radius.  
Note, that the mass within the virial radius estimated this way is simply an approximation, and it is expected to be uncertain given the unknown behaviour of the unmodeled mass beyond the region containing the lensing constraints. To account for the uncertainty due to the unknown profile in the outer regions, and to errors in the photometric redshitfs, we fit the two models bracketing the $1-\sigma$ regions from the 100 models described at the end of section \ref{sec_math}. The model in the upper $1-\sigma$ limit is assigned the shallower  $\beta=2$  slope which will correspond to the upper limit of the extrapolated mass. The model in the lower $1-\sigma$ limit is assigned the steeper  $\beta=3$ slope, and it will correspond to the lower limit of the extrapolated mass. Finally, the two gNFW models are added up together to account for the total mass. The resulting interval is shown as an orange shaded region in Figure \ref{Fig_TotalMass}. The upper limit corresponds to the upper 1-$\sigma$ model (among the 100 lens models) extrapolated with $\beta=2$ and the lower limit corresponds to the lower 1-$\sigma$ model (among the 100 lens models) extrapolated with $\beta=3$. The solid black line in Figure \ref{Fig_TotalMass} shows the resulting extrapolation for the driver model. For comparison, the black dashed line shows the double gNFW fit to the simulated cluster. Note how the driver model follows closely the simulation below $\approx 300$ kpc. Above $\approx 300$ kpc, the departure is mostly due to the memory effect at large radii, that is, the fact that the lens model is unconstrained beyond some radius and the solution recovers the initial condition of the optimization (small random values). The same departure is observed when we select the centre of the profile as the point in between the two subgroups and we skip the fit to the gNFW profiles. In this case (solid blue line in Figure \ref{Fig_TotalMass}) the departure is observed at $\approx 400$ kpc instead. When the centre of the profile is in between the two subgroups, the integrated mass of simulated cluster (blue dashed line) converges to the mass of the double gNFW profile at radius of $\approx 1$ Mpc. The light blue line marks the mass enclosed in a sphere with 200 times the critical density. The intersection with this curve marks the virial radius and virial mass. Based on the uncertainty of the extrapolation, and the dispersion due to photometric redshifts and randomness of the initial condition in the optimization,  we can estimate the mass at the virial radius from the intersection with the light blue line, $M_{200c} = 1.08^{+0.65}_{-0.12} \times 10^{15}\,M_{\odot}$. The extrapolation of the double gNFW model seems to imply that the lens model has a smaller virial mass than the simulated cluster. This tentative conclusion will need the JWST images, and spectra, to be confirmed by extending the range in radius of the lensing constraints.  The question of the total mass of El Gordo will be only settled with new data. JWST will provide new strong lensing constraints (for instance high-z galaxies extending farther away from the cluster central region), as well as weak lensing data which is scarce in current data given the high redshift of the cluster.

\subsection{The contribution from the gas and filamentary structure around the SE BCG}\label{Sect_Filament}
\begin{figure}
   \centering
   \plotone{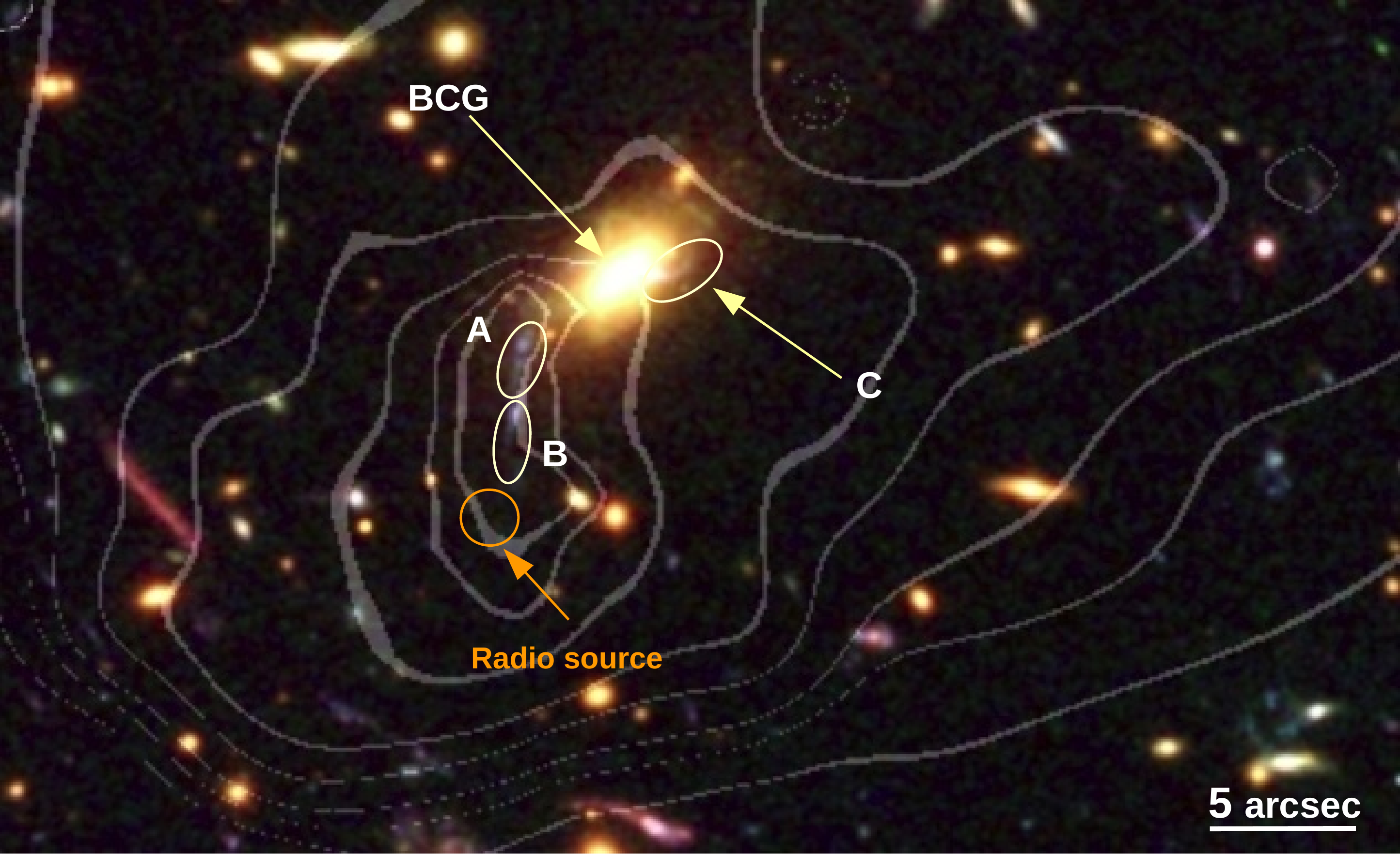}
      \caption{UV features in El Gordo. in The contours are the X-ray emission observed by Chandra. 
               Features A,B and C are visible in the bluer HST bands and are marked with yellow ellipses. 
               The orange circle marks the position of a compact radio source in Lindner et al. (2014). 
              }
         \label{Fig_CoolingFlow}
   \end{figure}

%----------------------------------------------------------------------------------------------------------
The position of the peak in the mass distribution of the full model is coincident with the peak in the X-ray emission. This coincidence is intriguing as it raises the possibility that the X-ray emitting gas is contributing substantially to the projected mass in this region of the lens plane. This is not all that surprising since clusters are expected to contain a baryon fraction representative of the cosmological value, and most of the baryons in massive clusters are in the form of a hot plasma. One expects the hot plasma to account for $\approx$ 10--15\% of the total mass in the entire cluster. Due to the more efficient cooling of baryons, this fraction increases towards the centre of clusters, so around the BCG one could expect an even larger contribution to the total mass from the hot plasma. No significant radio emission is observed near this position, except for a compact radio source named U7 in Figure 16 of \cite{Lindner2014}, and at 75 kpc SE from the BCG. This radio source is spatially coincident with the peak of the X-ray emission (see Figure \ref{Fig_CoolingFlow}). Based on X-ray data, \cite{Menanteau2012} constraints the electron density to values between 0.023 cm$^{-3}$ and 0.045  cm$^{-3}$ within a region of diameter $\approx 170$ kpc (or $\approx 22$''). Based on this electron density, the gas surface mass density (projected on 170 kpc along the line of sight) is then $\Sigma_{gas}\approx 100-200 {\rm M}_{\odot}$ pc$^{-2}$, which should be compared with the critical surface mass density of $\Sigma_{crit} = 2800 {\rm M}_{\odot}$ pc$^{-2}$ for a source at $z_s=3$ (note that the critical surface mass density is projected along a much longer line of sight). 
The contribution from the gas to the convergence, $\kappa$ (where $\kappa$ is defined as the ratio between the surface mass density and the critical surface mass density at the given lens and source redshifts), projected along this relatively small interval of 170 kpc is then 0.035-0.07. At the peak of the X-ray emission, and projecting over larger distances, the gas can easily contribute over 0.1 to $\kappa$. This may be sufficient to explain the correlation between the total mass peak and the X-ray emission shown in Figure~\ref{Fig_Mass_Xray}. 

%%----------------------------------------------------------------- 
%%-----------------------------------------------------------------

Interestingly, as discussed earlier, the peak of the X-ray emission coincides with blue features observed in the UV-optical bands (A and B in Figure~\ref{Fig_CoolingFlow}). Based on the driver model, if these two features are multiply lensed objects they need to be at a redshift above $z=1.8$. However, at this redshift, the predicted images would not form radially oriented arcs in our models, but rather tangential arcs. Radially oriented arcs at this position in the lens plane appear for redshifts $z\sim 2$ or above, but the lens model can not reproduce arcs with a morphology similar to the observed ones. For redshifts between $z=4.5$ and $z=5.5$, counterimages for the arcs are expected at positions compatible with the three arclets of system 4. This correspondence explains the original classification of these features as part of system 4, but the morphology of the predicted images differ significantly from the observation, making this possibility unlikely.
The two features are at distances of $\approx 35$ and $\approx 50$ kpc respectively from the centre of the BCG. 
The tight correlation shown in Figure~\ref{Fig_CoolingFlow} between these two features and the offset peak of the X-ray emission suggests that these two features may be the optical counterpart of a cooling flow (alternatively they could be associated with a jet emitting in UV-optical and X-rays). 
A well studied case that could serve as a similar example is the nearby cluster Abell 2597 (z=0.0821) where an arc-like feature correlates also very well with the peak in the X-ray emission. In \cite{Tremblay2012} they study this cluster combining data from X-ray, UV/optical, NIR and radio observations. They find evidence for a cooling flow in the X-ray band and filamentary features in the FUV and optical bands that resemble the blue features shown in Figure~\ref{Fig_CoolingFlow}, which could be associated with precipitation of the gas \citep{Voit2015}. If cold and dense gas is responsible for this feature, this could explain the apparent correlation between the X-ray and mass discussed in the previous subsection.

%%----------------------------------------------------------------- 

\section{Conclusions}\label{sect_concl}
%%%%%%%%%%%%%%%%%%%%%%%%%%%%%%%%%%%%%%%%%
We derive a new lens model and mass estimate for the El Gordo cluster using data from the RELICS program. This cluster is of special interest since earlier estimates suggest that it is the most massive cluster at redshift $z\sim0.9$, and hence it can be used to constrain the cosmological model, for instance by using extreme value statistics \citep{Harrison2012}. Understanding the uncertainties in the mass estimate emerging from the uncertainties in the lens model is of pivotal importance. In this work we present the first free-form model of this cluster that does not rely on assumptions about the mass distribution that, if they are erroneous, could bias the mass estimation. Free-form models offer a complementary cross-check with parametric models and are of particular interest in highly irregular clusters, like El Gordo, due to the lack of assumptions about the underlying distribution of dark matter.
We first derive a robust model (nicknamed the driver model) based on a reliable subsample of lensed galaxies. Using the driver model, we unveil new strongly lensed system candidates and infer their redshifts. With the full set of lens systems we derive an alternative model (or full model) for the mass distribution. Both models are similar to each other, but small differences can be identified, specially in the SE sector of the cluster. We explicitly compare our models with the one derived by \cite{Cerny2018} using the same RELICS data, but a different sample of lensed galaxies (although with substantial overlap between our sample and theirs). We find that our lens model predicts wider critical curves, but the integrated mass as a function of aperture is consistent with the model of \cite{Cerny2018}, thus suggesting that the mass estimate obtained with earlier parametric methods may not be affected by erroneous choices in the parameterization. Our new model predicts also nearly twice the lensing efficiency above a given magnification factor (at large magnifications).  

We explore uncertainties in the lens model due to errors in the assumed redshifts. Adopting an uncertainty of 0.5 in the redshift of all systems, we produce a set of 100 models where we vary the redshift with this uncertainty. We find that the mass within 300 kpc of each subgroup varies by relatively small amounts due to this uncertainty in redshift. By fitting our full lens model to an NFW profile, and extrapolating up to $R_{200c}$, we find a mass $M_{200c}=(1.08^{+0.65}_{-0.12})\times10^{15}$M$_{\odot}$, where the uncertainty comes mostly from the poorly constrained scale radius (and the external slope $\beta$).  Due to the highly non-symmetric nature of the cluster, the extrapolation to the virial radius is not as reliable as it would be in more symmetric clusters. Hence, this mass estimate should be taken with caution. Improved identification and redshift determinations of the lensed galaxies, together with the addition of dense weak lensing measurements will be needed to improve the mass estimate up to the virial radius. Our mass estimate is smaller than previous estimates. If confirmed, this would relax the tension with standard LCDM models that predict that clusters with masses above $M_{200\rho}=1.7\times10^{15}$M$_{\odot}$ and at this redshift should be very rare (this cluster was serendipitously found in a relatively small 2\% portion of the sky). We test the accuracy of the profile extrapolation using an N-body simulation that matches most of the observed features in El Gordo, and conclude that our mass estimate may still be underestimating the real mass due to systematics in the extrapolation of the profile. A combination of future strong and weak lensing data (for instance with JWST which will be able to detect the high redshift background galaxies needed for the weak lensing analysis of this high-z cluster) should allow for a better constrain of the total mass of this cluster.  

We find evidence for the lens model being sensitive to the gas mass. In particular, we find that the peak of the smooth component of the mass distribution in the full lens model agrees well with the peak of the X-ray emission (which is offset with respect to the nearby BCG). We discuss the possibility that two features at the location of these peaks, which are observed in the optical-UV bands, and interpreted in the past as background lensed galaxies, are instead the optical counterpart of a cooling flow, or a precipitation mechanism from the hot plasma. 

New lens models will be valuable when El Gordo is observed as part of of Cycle 1 of JWST (as part of the JWST Medium Deep Fields program, P.I; R. Windhorst). New arc systems, including several at high redshift are expected to be discovered with the new JWST data. Our lens model can be used to identify new strongly lensed system candidates, as well as to estimate their redshifts.  

\acknowledgements
%%%%%%%%%%%%%%%%%%%%%%%%%
J.M.D. acknowledges the support of projects AYA2015-64508-P (MINECO/FEDER, UE), funded by the Ministerio de Economia y Competitividad and PGC2018-101814-B-100 (MCIU/AEI/MINECO/FEDER, UE) Ministerio de Ciencia, Investigaci\'on y Universidades. This project was funded by the Agencia Estatal de Investigaci\'on, Unidad de Excelencia Mar\'ia de Maeztu, ref. MDM-2017-0765. 
This work was funded by NASA JWST Interdisciplinary Scientist grants NAG5-12460, NNX14AN10G, and 80NSSC18K0200 to RAW from GSFC. We would like to thank Harald Ebeling for making  the code {\small ASMOOTH} \citep{Ebeling2006} available. J.M.D. acknowledges the hospitality of the Physics Department at the University of Pennsylvania for hosting him during the preparation of this work. This work is based on observations made with the NASA/ESA {\it Hubble Space Telescope} and operated by the Association of Universities for Research in Astronomy, Inc. under NASA contract NAS 5-2655. Part of the data for this study is retrieved from the Mikulski Archive for Space Telescope (MAST). The authors would like to thank the RELICS team for making the reduced data set available to the community. The scientific results reported in this article are based in part on data obtained from the Chandra Data Archive \footnote{ivo://ADS/Sa.CXO\#obs/12258}\footnote{ivo://ADS/Sa.CXO\#obs/14022}\footnote{ivo://ADS/Sa.CXO\#obs/14023}. 
%\end{acknowledgements}

%%%%%%%%%%%%%%%%%%%%%%%%%%%%%%%%%%%%%%%%%%%%%%%%%%%%%%%%%%%%%%%%%%%%%%%%%%%%%%%%%%%%

\newpage

\appendix

\section{Compilation of arc positions}
This appendix presents the sample of secure and likely lensed multiple images detected behind El Gordo using the updated imaging from the RELICS program. Table~\ref{tab_arcs} lists the complete sample of images and their redshifts assigning IDs to each of them. 

The first column shows system ID following the original notation of \cite{Zitrin2013} (ID1.ID2.ID3 = System.Image.Knot) and ranks (A, B and C) . 
Systems 1-4 were initially presented by \cite{Zitrin2013}. IDs marked with bold face are new systems presented in this work.  
Photometric redshifts are given in column $z_{\rm phot}$. 
The systems having spectroscopic redshift are marked with bold face. 
Redshifts predicted by the lens model are given in column  $z_{\rm model}$. 
The redshift used to reconstruct the lens are given in column  $z_{\rm used}$. 

The column labeled Rank shows the quality of the system. Systems marked with rank A are very reliable and are used to derive the {\it driver} model. 
Systems marked with B are used to derive (together with systems having rank A) the {\it full} model. 
Systems marked with C are less reliable, but still highly consistent with the driver lens model.  
In the last column 1, 2 and 3 refer to previous work, where these systems are defined.  
Z13 stands for \cite{Zitrin2013} while C18 for \cite{Cerny2018}. The number in parenthesis next to the reference indicates the system ID in the corresponding publication. 

%%%%%%%%%%%%%%%%%%%%%%%%%%%%%%%%%%%%%%%%%%%%%%%%%%%%%%%%%%%%%%%%%%%%%%%%%%%%%%%%%%%%%%%%%%%%%%

\begin{table*}
  \begin{minipage}{165mm}                                               
    \caption{Full strong lensing data set. See text for description of the columns. 
             For the photometric redshifts we indicate the range of redshifts (from multiple images) after excluding extreme values. Systems marked in bold face are newly identified systems.}
 \label{tab_arcs}
 \begin{tabular}{|cccccccc|}   
 \hline
%--------------------------------------------------------------------------------------------------------------------------------------------------
   KnotID   &       RA       &     DEC   & $z_{\rm phot}$ & $z_{\rm model}$ &  $z_{\rm used}$& Rank & Comments \\
%---------------------------------------------------------------------------------------------------------------------------------------------------
 \hline
 1.1.1   &  1 02 53.275  & -49 15 16.49   &            &   3     & 3    &  A   & Z13(1)   \\    
 1.2.1   &  1 02 52.819  & -49 15 18.29   &   2.93     &         &      &  A   &    \\
 1.3.1   &  1 02 55.411  & -49 14 59.90   &            &         &      &  A   &    \\
 1.1.2   &  1 02 53.340  & -49 15 16.36   &            &         &      &  A   &    \\
 1.2.2   &  1 02 52.763  & -49 15 18.70   &   2.8      &         &      &  A   &    \\
 1.3.2   &  1 02 55.391  & -49 15 00.33   &   2.91     &         &      &  A   &    \\
 1.1.3   &  1 02 53.480  & -49 15 16.01   &            &         &      &  A   &    \\
 1.2.3   &  1 02 52.600  & -49 15 19.68   &            &         &      &  A   &    \\
 1.3.3   &  1 02 55.320  & -49 15 01.18   &   3.26     &         &      &  A   &    \\
 \hline
 2.1.1   &  1 02 55.828  & -49 15 52.37   &   3.21     &    3.3  & 3.3  &  A   &  Z13(2)    \\
 2.2.1   &  1 02 56.749  & -49 15 46.01   &   3.39     &         &      &  A   &    \\
 2.3.1   &  1 02 54.429  & -49 16 04.63   &   3.3      &         &      &  A   &    \\
 2.1.2   &  1 02 55.671  & -49 15 53.54   &            &         &      &  A   &    \\
 2.2.2   &  1 02 56.885  & -49 15 45.17   &   3.27     &         &      &  A   &    \\
 2.3.2   &  1 02 54.456  & -49 16 04.00   &   2.9      &         &      &  A   &    \\
 2.1.3   &  1 02 55.983  & -49 15 51.24   &            &         &      &  A   &    \\
 2.2.3   &  1 02 56.573  & -49 15 47.06   &            &         &      &  A   &    \\ 
 2.3.3   &  1 02 54.383  & -49 16 04.61   &            &         &      &  A   &    \\
 \hline
 3.1.1   &  1 02 56.257  & -49 15 07.03   &            &   4.4   & 4.4  &  A   &  Z13(3)    \\
 3.2.1   &  1 02 54.751  & -49 15 19.54   &            &         &      &  A   &    \\
 3.3.1   &  1 02 51.536  & -49 15 38.47   &  4.54      &         &      &  A   &    \\
 \hline
 4.1.1   &  1 02 59.986  & -49 15 49.54   &  3.98      &   3.2   & 4    &  A   &  Z13(4)    \\
 4.2.1   &  1 02 55.362  & -49 16 26.09   &  4.0       &         &      &  A   &    \\
 4.3.1   &  1 02 56.599  & -49 16 08.45   &            &         &      &  A   &    \\
 \hline
 5.1.1  &  1 02 54.539  & -49 14 58.60   &  2.4       &     2.8 & 2.8  &  A   & C18(13)   \\
 5.2.1  &  1 02 53.230  & -49 15 07.11   &           &          &      &  A   &    \\   
 5.3.1  &  1 02 51.803  & -49 15 17.05   &  2.2,2.5   &         &      &  A   &    \\ 
 \hline
{\bf 6.1.1} & 1 02 55.484  & -49 15 05.04   &            &   4.3   & 4.3  &  B   &    \\
{\bf 6.2.1} & 1 02 55.067  & -49 15 09.84   &            &   4.3   &      &  B   &    \\
{\bf 6.3.1} & 1 02 51.242  & -49 15 37.08   &  4.3,4.5   &   4.3   &      &  C   &    \\
{\bf 6.1.2} & 1 02 55.330  & -49 15 05.70   &            &         &      &  B   &    \\     
{\bf 6.2.2} & 1 02 55.134  & -49 15 07.80   &            &         &      &  B   &    \\
{\bf 6.3.2} & 1 02 51.193  & -49 15 37.08   &            &         &      &  C   &    \\
 \hline
 7.1.1   &  1 02 55.477  & -49 16 07.32   &  4.53      &   4.5   &      &  B   & Z13   \\
 7.2.1   &  1 02 54.927  & -49 16 14.85   &            &         &      &  B   &    \\   
 7.3.1   &  1 02 59.321  & -49 15 44.52   &  4.8       &         &      &  C   &    \\     
 \hline
 8.1.1   &  1 02 55.836  & -49 16 07.56   &  3.55      &   4     & 3.5  &  D   & Z13   \\
 8.2.1   &  1 02 55.211  & -49 16 16.10   &            &         &      &  D   &    \\ 
 \hline
 9.1.1   &  1 02 56.288  & -49 16 07.90   &  2.72      &   3     & 2.9  &  B   & Z13   \\
 9.2.1   &  1 02 55.641  & -49 16 17.54   &  2.26      &         &      &  B   &    \\ 
 9.3.1   &  1 02 59.043  & -49 15 53.35   &  2.32      &         &      &  C   &    \\
 \hline
{\bf 10.1.1} & 1 02 55.784  & -49 15 13.91 &  5.1       &    5.15 & 5.1  &  B   &    \\
{\bf 10.2.1} & 1 02 55.558  & -49 15 15.99 &            &         &      &  B   &    \\ 
{\bf 10.3.1} & 1 02 51.772  & -49 15 44.75 &            &         &      &  C   &    \\
 \hline
 11.1.1   &  1 02 59.612  & -49 16 26.61   &  2.19      &   3.1   & 2.2  &  B   & Z13(5)   \\
 11.2.1   &  1 02 59.467  & -49 16 27.99   &            &         &      &  B   &    \\  
{\bf 11.3.1} & 1 02 57.774 & -49 16 39.10  &            &         &      &  B   &    \\
 \hline
 12.1.1  &  1 02 54.571  & -49 14 54.16   &  3.36      &     3   & 3    &  B   & C18(14)   \\
{\bf 12.2.1} & 1 02 53.021 & -49 15 04.94 &            &         &      &  B   &    \\
{\bf 12.3.1} & 1 02 51.782 & -49 15 14.38 &  2.8       &         &      &  B   &    \\
 \hline
{\bf 13.1.1} & 1 02 59.884  & -49 16 30.53 &            &   2.4  & 3    &  B   &    \\
{\bf 13.2.1} & 1 02 59.719  & -49 16 32.59 &            &        &      &  B   &    \\ 
 \hline

 \end{tabular}  
 \end{minipage}
\end{table*}

\setcounter{table}{0}
 \begin{table*}
    \begin{minipage}{165mm}                                               
    \caption{Continuation}
\begin{tabular}{|cccccccc|}   
 \hline
%--------------------------------------------------------------------------------------------------------------------------------------------------
   KnotID   &       RA       &     DEC        & $z_{phot}$   & $z_{\rm model}$ &  $z_{\rm used}$  & Rank & Comments \\
%---------------------------------------------------------------------------------------------------------------------------------------------------
 \hline
{\bf 14.1.1} & 1 03 00.135 & -49 15 46.29   &  2.74     &     4   & 4    &  B   &    \\
{\bf 14.2.1} & 1 02 55.161 & -49 16 23.07   &           &     4   &      &  B   &    \\
{\bf 14.3.1} & 1 02 56.331 & -49 16 08.55   &           &         &      &  B   &    \\ 
 \hline
 15.1.1  &  1 02 58.512  & -49 16 37.00   &  2.7       &  2.65   & 2.7  &  B   &  C18(5)  \\
 15.2.1  &  1 02 58.736  & -49 16 35.71   &  2.8       &         &      &  B   &    \\
 15.3.1  &  1 03 00.100  & -49 16 21.12   &            &         &      &  C   &    \\
 \hline
{\bf 16.1.1} & 1 02 58.017 & -49 15 33.48 &            &   4.3   & 4.1  &  B   &    \\ 
{\bf 16.2.1} & 1 02 55.237 & -49 15 53.35 &            &         &      &  B   &    \\ 
{\bf 16.3.1} & 1 02 53.719 & -49 16 01.99 &  4.13      &         &      &  B   &    \\
 \hline
{\bf 17.1.1} &  1 02 55.781  & -49 15 00.23   &  4.2   &   5     & 5    &  B   &    \\
{\bf 17.2.1} &  1 02 54.274  & -49 15 12.48   &        &         &      &  B   &    \\
{\bf 17.3.1} &  1 02 50.950  & -49 15 33.57   &  4.4   &         &      &  B   &    \\
 \hline
{\bf 18.1.1} &  1 02 57.018  & -49 15 47.45   &        &   3.4   & 3.3  &  C   &    \\
{\bf 18.2.1} &  1 02 55.784  & -49 15 56.22   &  3.27  &         &      &  C   &    \\
{\bf 18.3.1} &  1 02 54.575  & -49 16 06.89   &        &         &      &  C   &    \\ 
 \hline
{\bf 19.1.1} &  1 02 52.709  & -49 15 51.82   &        &  4.5    & 5    &  C   &    \\
{\bf 19.2.1} &  1 02 55.275  & -49 15 33.43   &        &         &      &  C   &    \\   
{\bf 19.3.1} &  1 02 56.886  & -49 15 21.16   &        &         &      &  C   &    \\ 
 \hline

 \end{tabular}
 \end{minipage}
\end{table*}

\newpage 

\section{Predicted arcs}
\begin{figure*}
   \centering
   \plotone{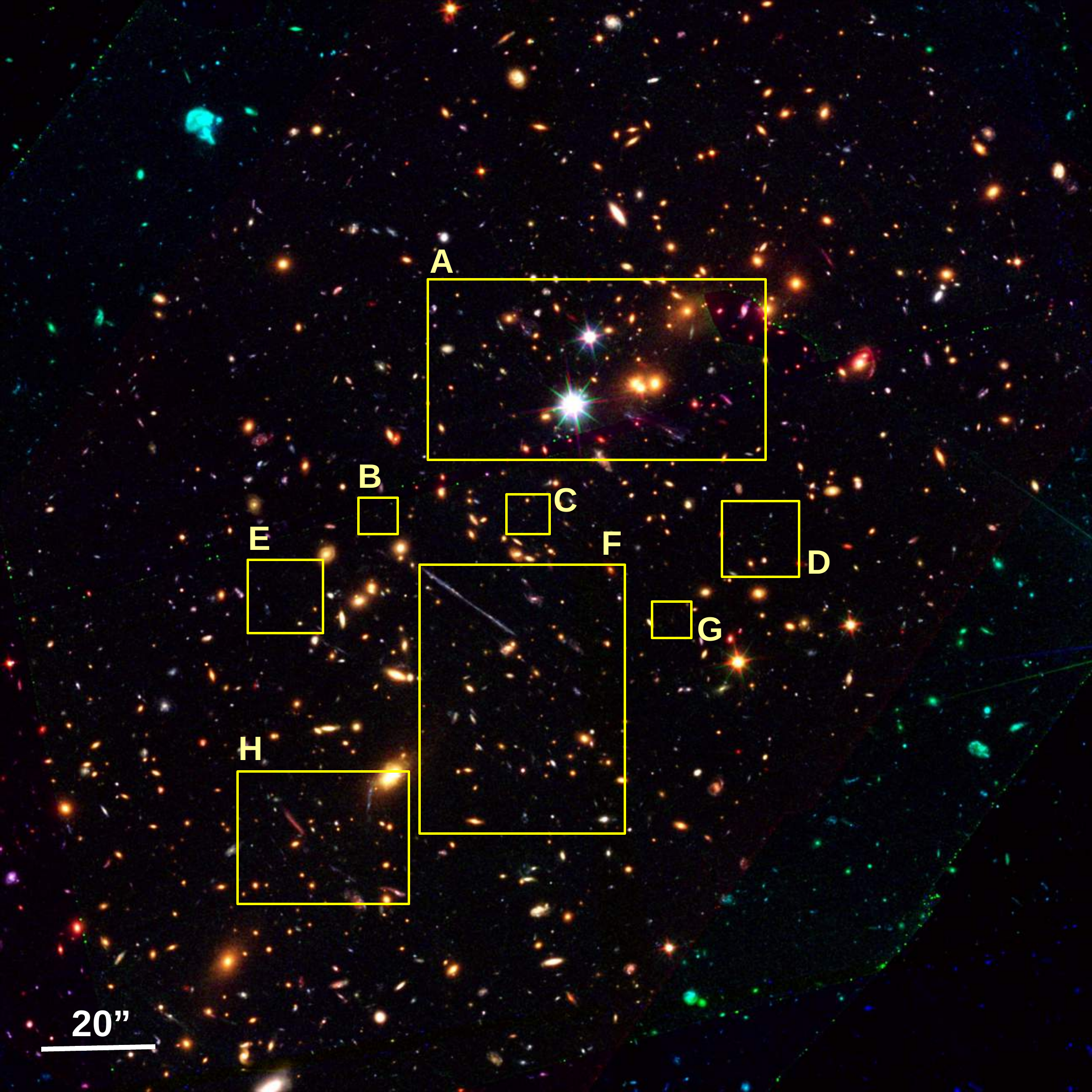}
      \caption{Rectangles mark the seven regions where the prediction made by the driver model is compared with the data. This prediction is shown in Figure \ref{Fig_Regions2Zoom2} below. 
              }
         \label{Fig_Regions2Zoom1}
   \end{figure*}

\begin{figure}
   \centering
   \plotone{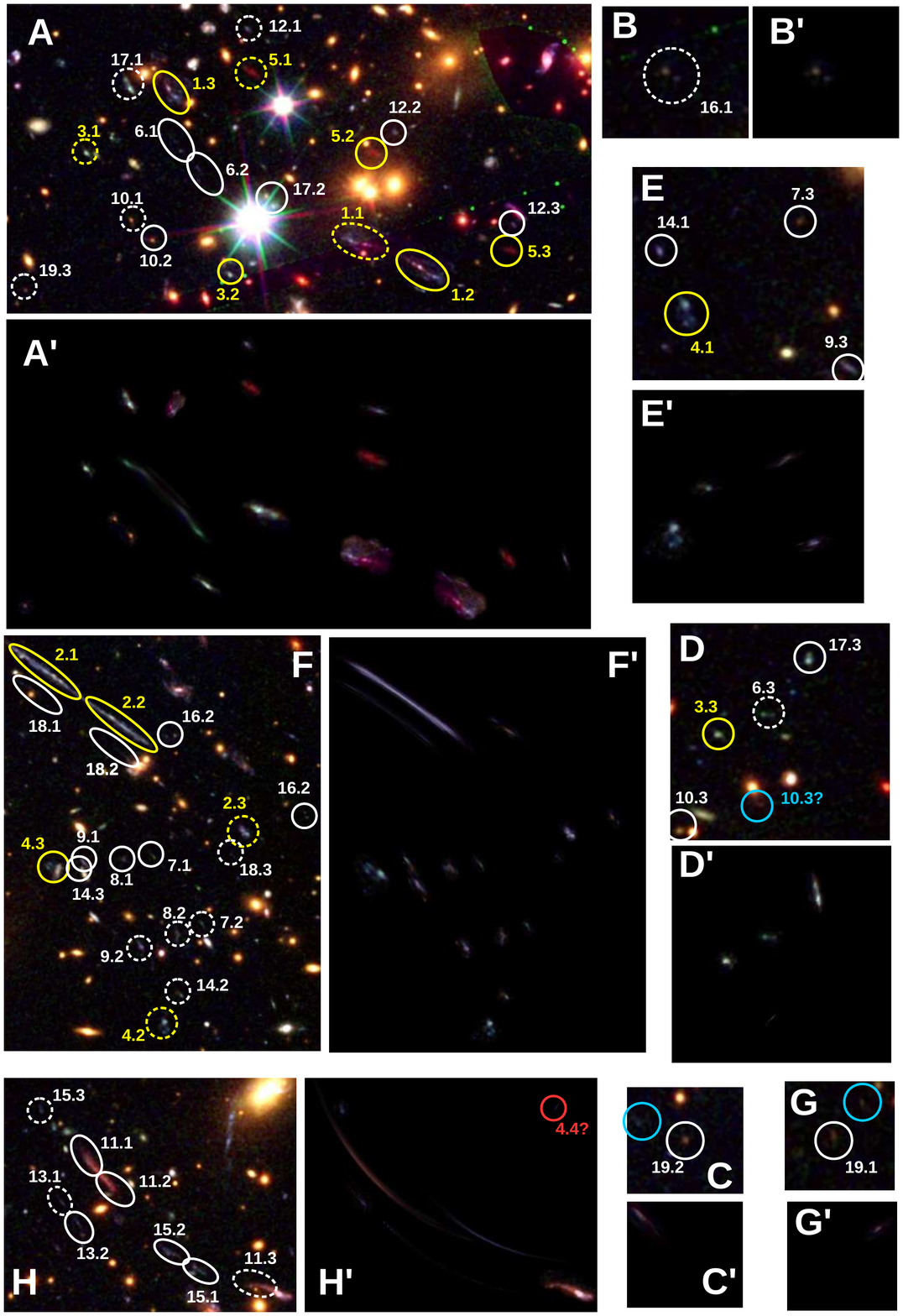}
      \caption{Zoomed regions shown in Figure \ref{Fig_Regions2Zoom1}. Panels with primed letters denote the images predicted by the driver model. Dashed line circles show the image that is delensed and later relensed. Solid line circles mark the images being predicted by the driver model. The systems used to derive the driver model are marked in yellow. The scale and reference points are the same for each pair of panels. 
              }
         \label{Fig_Regions2Zoom2}
   \end{figure}

%%%%%%%%%%%%%%%%%%%%%%%%%
In this section we show the position and morphology of the images predicted by the driver model. In Figure \ref{Fig_Regions2Zoom1} we show the entire field of view and define 8 regions that contain the multiply lensed images. These regions are later shown separately in Figure \ref{Fig_Regions2Zoom2} where we represent both the data (in panels labeled, A, B, C, D, E F, G, and H) and the prediction made by the driver model (panels labeled with the corresponding primed letters). The orientation, scale and reference position is the same for each pair of panels. For each multiply lensed system, one of the counterimages is chosen (dashed circles), delensed into the source plane, and relensed into the image plane (solid line circles). We remind the reader that the driver model is derived from just 5 multiple lensed systems shown as yellow circles and marked with yellow labels in Figure \ref{Fig_Regions2Zoom2}. Hence, deviations between the data and the prediction made by the model are expected to be larger for the systems (like system 10, 11 , and 13)) that where not used to derived the driver lens model (i.e, white circles in Figure \ref{Fig_Regions2Zoom2}). Note that when systems are not used as constraints in the model, deviations in position similar to those shown in  Figure \ref{Fig_Regions2Zoom2} are expected, specially in regions of the lens plane where constraints are more scarce. In some cases, we mark with blue circles possible alternative candidates. A red circle is used in panel H' to mark the position of an arclet predicted by the model but not observed which would correspond to system 4. However, this arclet falls within the BCG and overlapping with the blue UV regions discussed in section \ref{Sect_Filament}, so it is possible that the arclet is merged with the UV regions. 
%%----------------------------------------------------------------- 
%%-----------------------------------------------------------------

\newpage 

%%----------------------------------------------------------------- 
%%-----------------------------------------------------------------

\newpage

%%%%%%%%%%%%%%%%%%%%%%%%%%%%%%%%%%%%%%%%%%%%%%%%%%%%%%%%%%%%%%%%%%%%%%%%%%%%%%%%%%%5

\bibliographystyle{aasjournal} %use aasjournal.bst
\bibliography{MyBiblio} % References in MyBiblio.bib run bibtex after latex/pdflatex 

\begin{thebibliography}{}
\expandafter\ifx\csname natexlab\endcsname\relax\def\natexlab#1{#1}\fi
\providecommand{\url}[1]{\href{#1}{#1}}
\providecommand{\dodoi}[1]{doi:~\href{http://doi.org/#1}{\nolinkurl{#1}}}
\providecommand{\doeprint}[1]{\href{http://ascl.net/#1}{\nolinkurl{http://ascl.net/#1}}}
\providecommand{\doarXiv}[1]{\href{https://arxiv.org/abs/#1}{\nolinkurl{https://arxiv.org/abs/#1}}}

\bibitem[{{Botteon} {et~al.}(2016){Botteon}, {Gastaldello}, {Brunetti}, \&
  {Kale}}]{Botteon2016}
{Botteon}, A., {Gastaldello}, F., {Brunetti}, G., \& {Kale}, R. 2016, \mnras,
  463, 1534, \dodoi{10.1093/mnras/stw2089}

\bibitem[{{Cerny} {et~al.}(2018){Cerny}, {Sharon}, {Andrade-Santos}, {Avila},
  {Brada{\v c}}, {Bradley}, {Carrasco}, {Coe}, {Czakon}, {Dawson}, {Frye},
  {Hoag}, {Huang}, {Johnson}, {Jones}, {Lam}, {Lovisari}, {Mainali}, {Oesch},
  {Ogaz}, {Past}, {Paterno-Mahler}, {Peterson}, {Riess}, {Rodney}, {Ryan},
  {Salmon}, {Sendra-Server}, {Stark}, {Strolger}, {Trenti}, {Umetsu},
  {Vulcani}, \& {Zitrin}}]{Cerny2018}
{Cerny}, C., {Sharon}, K., {Andrade-Santos}, F., {et~al.} 2018, \apj, 859, 159,
  \dodoi{10.3847/1538-4357/aabe7b}

\bibitem[{{Chan} {et~al.}(2020){Chan}, {Broadhurst}, {Lim}, {Wong}, {Diego}, \&
  {Coe}}]{Chan2020}
{Chan}, B. M.~Y., {Broadhurst}, T., {Lim}, J., {et~al.} 2020, \apj, 888, 35,
  \dodoi{10.3847/1538-4357/ab44a4}

\bibitem[{{Coe} {et~al.}(2019){Coe}, {Salmon}, {Bradac}, {Bradley}, {Sharon},
  {Zitrin}, {Acebron}, {Cerny}, {Cibirka}, {Strait}, {Paterno-Mahler},
  {Mahler}, {Avila}, {Ogaz}, {Huang}, {Pelliccia}, {Stark}, {Mainali}, {Oesch},
  {Trenti}, {Carrasco}, {Dawson}, {Rodney}, {Strolger}, {Riess}, {Jones},
  {Frye}, {Czakon}, {Umetsu}, {Vulcani}, {Graur}, {Jha}, {Graham}, {Molino},
  {Nonino}, {Hjorth}, {Selsing}, {Christensen}, {Kikuchihara}, {Ouchi},
  {Oguri}, {Welch}, {Lemaux}, {Andrade-Santos}, {Hoag}, {Johnson}, {Peterson},
  {Past}, {Fox}, {Agulli}, {Livermore}, {Ryan}, {Lam}, {Sendra-Server}, {Toft},
  {Lovisari}, \& {Su}}]{Coe2019}
{Coe}, D., {Salmon}, B., {Bradac}, M., {et~al.} 2019, arXiv e-prints.
\newblock \doarXiv{1903.02002}

\bibitem[{{Diego}(2018)}]{Diego2019}
{Diego}, J.~M. 2018, arXiv e-prints.
\newblock \doarXiv{1806.04668}

\bibitem[{{Diego} {et~al.}(2005a){Diego}, {Protopapas}, {Sandvik}, \&
  {Tegmark}}]{Diego2005}
{Diego}, J.~M., {Protopapas}, P., {Sandvik}, H.~B., \& {Tegmark}, M. 2005a,
  \mnras, 360, 477, \dodoi{10.1111/j.1365-2966.2005.09021.x}

\bibitem[{{Diego} {et~al.}(2007){Diego}, {Tegmark}, {Protopapas}, \&
  {Sandvik}}]{Diego2007}
{Diego}, J.~M., {Tegmark}, M., {Protopapas}, P., \& {Sandvik}, H.~B. 2007,
  \mnras, 375, 958, \dodoi{10.1111/j.1365-2966.2007.11380.x}

\bibitem[{{Diego} {et~al.}(2016){Diego}, {Broadhurst}, {Chen}, {Lim}, {Zitrin},
  {Chan}, {Coe}, {Ford}, {Lam}, \& {Zheng}}]{Diego2016}
{Diego}, J.~M., {Broadhurst}, T., {Chen}, C., {et~al.} 2016, \mnras, 456, 356,
  \dodoi{10.1093/mnras/stv2638}

\bibitem[{{Diego} {et~al.}(2018){Diego}, {Kaiser}, {Broadhurst}, {Kelly},
  {Rodney}, {Morishita}, {Oguri}, {Ross}, {Zitrin}, {Jauzac}, {Richard},
  {Williams}, {Vega-Ferrero}, {Frye}, \& {Filippenko}}]{Diego2018}
{Diego}, J.~M., {Kaiser}, N., {Broadhurst}, T., {et~al.} 2018, \apj, 857, 25,
  \dodoi{10.3847/1538-4357/aab617}

\bibitem[{{Donnert}(2014)}]{Donnert2014}
{Donnert}, J.~M.~F. 2014, \mnras, 438, 1971, \dodoi{10.1093/mnras/stt2291}

\bibitem[{{Ebeling} {et~al.}(2006){Ebeling}, {White}, \&
  {Rangarajan}}]{Ebeling2006}
{Ebeling}, H., {White}, D.~A., \& {Rangarajan}, F.~V.~N. 2006, \mnras, 368, 65,
  \dodoi{10.1111/j.1365-2966.2006.10135.x}

\bibitem[{{Hallman} \& {Markevitch}(2004)}]{Hallman2004}
{Hallman}, E.~J., \& {Markevitch}, M. 2004, \apjl, 610, L81,
  \dodoi{10.1086/423449}

\bibitem[{{Harrison} \& {Coles}(2012)}]{Harrison2012}
{Harrison}, I., \& {Coles}, P. 2012, \mnras, 421, L19,
  \dodoi{10.1111/j.1745-3933.2011.01198.x}

\bibitem[{{Jee} {et~al.}(2014){Jee}, {Hughes}, {Menanteau}, {Sif{\'o}n},
  {Mandelbaum}, {Barrientos}, {Infante}, \& {Ng}}]{Jee2014}
{Jee}, M.~J., {Hughes}, J.~P., {Menanteau}, F., {et~al.} 2014, \apj, 785, 20,
  \dodoi{10.1088/0004-637X/785/1/20}

\bibitem[{{Kelly} {et~al.}(2018){Kelly}, {Diego}, {Rodney}, {Kaiser},
  {Broadhurst}, {Zitrin}, {Treu}, {P{\'e}rez-Gonz{\'a}lez}, {Morishita},
  {Jauzac}, {Selsing}, {Oguri}, {Pueyo}, {Ross}, {Filippenko}, {Smith},
  {Hjorth}, {Cenko}, {Wang}, {Howell}, {Richard}, {Frye}, {Jha}, {Foley},
  {Norman}, {Bradac}, {Zheng}, {Brammer}, {Benito}, {Cava}, {Christensen}, {de
  Mink}, {Graur}, {Grillo}, {Kawamata}, {Kneib}, {Matheson}, {McCully},
  {Nonino}, {P{\'e}rez-Fournon}, {Riess}, {Rosati}, {Schmidt}, {Sharon}, \&
  {Weiner}}]{Kelly2018}
{Kelly}, P.~L., {Diego}, J.~M., {Rodney}, S., {et~al.} 2018, Nature Astronomy,
  2, 334, \dodoi{10.1038/s41550-018-0430-3}

\bibitem[{{Lindner} {et~al.}(2014){Lindner}, {Baker}, {Hughes}, {Battaglia},
  {Gupta}, {Knowles}, {Marriage}, {Menanteau}, {Moodley}, {Reese}, \&
  {Srianand}}]{Lindner2014}
{Lindner}, R.~R., {Baker}, A.~J., {Hughes}, J.~P., {et~al.} 2014, \apj, 786,
  49, \dodoi{10.1088/0004-637X/786/1/49}

\bibitem[{{Mathis} {et~al.}(2005){Mathis}, {Lavaux}, {Diego}, \&
  {Silk}}]{Mathis2005}
{Mathis}, H., {Lavaux}, G., {Diego}, J.~M., \& {Silk}, J. 2005, \mnras, 357,
  801, \dodoi{10.1111/j.1365-2966.2004.08589.x}

\bibitem[{{Menanteau} {et~al.}(2012){Menanteau}, {Hughes}, {Sif{\'o}n},
  {Hilton}, {Gonz{\'a}lez}, {Infante}, {Barrientos}, {Baker}, {Bond}, {Das},
  {Devlin}, {Dunkley}, {Hajian}, {Hincks}, {Kosowsky}, {Marsden}, {Marriage},
  {Moodley}, {Niemack}, {Nolta}, {Page}, {Reese}, {Sehgal}, {Sievers},
  {Spergel}, {Staggs}, \& {Wollack}}]{Menanteau2012}
{Menanteau}, F., {Hughes}, J.~P., {Sif{\'o}n}, C., {et~al.} 2012, \apj, 748, 7,
  \dodoi{10.1088/0004-637X/748/1/7}

\bibitem[{{Molnar} \& {Broadhurst}(2015)}]{Molnar2015}
{Molnar}, S.~M., \& {Broadhurst}, T. 2015, \apj, 800, 37,
  \dodoi{10.1088/0004-637X/800/1/37}

\bibitem[{{Molnar} \& {Broadhurst}(2018)}]{Molnar2018}
---. 2018, \apj, 862, 112, \dodoi{10.3847/1538-4357/aad04c}

\bibitem[{{Ng} {et~al.}(2015){Ng}, {Dawson}, {Wittman}, {Jee}, {Hughes},
  {Menanteau}, \& {Sif{\'o}n}}]{Ng2015}
{Ng}, K.~Y., {Dawson}, W.~A., {Wittman}, D., {et~al.} 2015, \mnras, 453, 1531,
  \dodoi{10.1093/mnras/stv1713}

\bibitem[{{Sendra} {et~al.}(2014){Sendra}, {Diego}, {Broadhurst}, \&
  {Lazkoz}}]{Sendra2014}
{Sendra}, I., {Diego}, J.~M., {Broadhurst}, T., \& {Lazkoz}, R. 2014, \mnras,
  437, 2642, \dodoi{10.1093/mnras/stt2076}

\bibitem[{{Tremblay} {et~al.}(2012){Tremblay}, {O'Dea}, {Baum}, {Clarke},
  {Sarazin}, {Bregman}, {Combes}, {Donahue}, {Edge}, {Fabian}, {Ferland},
  {McNamara}, {Mittal}, {Oonk}, {Quillen}, {Russell}, {Sanders}, {Salom{\'e}},
  {Voit}, {Wilman}, \& {Wise}}]{Tremblay2012}
{Tremblay}, G.~R., {O'Dea}, C.~P., {Baum}, S.~A., {et~al.} 2012, \mnras, 424,
  1042, \dodoi{10.1111/j.1365-2966.2012.21278.x}

\bibitem[{{Vega-Ferrero} {et~al.}(2019){Vega-Ferrero}, {Diego}, \&
  {Bernstein}}]{Vega-Ferrero2019}
{Vega-Ferrero}, J., {Diego}, J.~M., \& {Bernstein}, G.~M. 2019, \mnras,
  \dodoi{10.1093/mnras/stz1217}

\bibitem[{{Voit} {et~al.}(2015){Voit}, {Donahue}, {Bryan}, \&
  {McDonald}}]{Voit2015}
{Voit}, G.~M., {Donahue}, M., {Bryan}, G.~L., \& {McDonald}, M. 2015, \nat,
  519, 203, \dodoi{10.1038/nature14167}

\bibitem[{{Waizmann} {et~al.}(2012){Waizmann}, {Ettori}, \&
  {Moscardini}}]{Waizmann2012}
{Waizmann}, J.~C., {Ettori}, S., \& {Moscardini}, L. 2012, \mnras, 420, 1754,
  \dodoi{10.1111/j.1365-2966.2011.20171.x}

\bibitem[{{Watson} {et~al.}(2014){Watson}, {Iliev}, {Diego}, {Gottl{\"o}ber},
  {Knebe}, {Mart{\'{\i}}nez-Gonz{\'a}lez}, \& {Yepes}}]{Watson2014}
{Watson}, W.~A., {Iliev}, I.~T., {Diego}, J.~M., {et~al.} 2014, \mnras, 437,
  3776, \dodoi{10.1093/mnras/stt2173}

\bibitem[{{Williamson} {et~al.}(2011){Williamson}, {Benson}, {High},
  {Vanderlinde}, {Ade}, {Aird}, {Andersson}, {Armstrong}, {Ashby}, {Bautz},
  {Bazin}, {Bertin}, {Bleem}, {Bonamente}, {Brodwin}, {Carlstrom}, {Chang},
  {Chapman}, {Clocchiatti}, {Crawford}, {Crites}, {de Haan}, {Desai}, {Dobbs},
  {Dudley}, {Fazio}, {Foley}, {Forman}, {Garmire}, {George}, {Gladders},
  {Gonzalez}, {Halverson}, {Holder}, {Holzapfel}, {Hoover}, {Hrubes}, {Jones},
  {Joy}, {Keisler}, {Knox}, {Lee}, {Leitch}, {Lueker}, {Luong-Van}, {Marrone},
  {McMahon}, {Mehl}, {Meyer}, {Mohr}, {Montroy}, {Murray}, {Padin}, {Plagge},
  {Pryke}, {Reichardt}, {Rest}, {Ruel}, {Ruhl}, {Saliwanchik}, {Saro},
  {Schaffer}, {Shaw}, {Shirokoff}, {Song}, {Spieler}, {Stalder}, {Stanford},
  {Staniszewski}, {Stark}, {Story}, {Stubbs}, {Vieira}, {Vikhlinin}, \&
  {Zenteno}}]{Williamson2011}
{Williamson}, R., {Benson}, B.~A., {High}, F.~W., {et~al.} 2011, \apj, 738,
  139, \dodoi{10.1088/0004-637X/738/2/139}

\bibitem[{{Windhorst} {et~al.}(2018){Windhorst}, {Timmes}, {Wyithe},
  {Alpaslan}, {Andrews}, {Coe}, {Diego}, {Dijkstra}, {Driver}, {Kelly}, \&
  {Kim}}]{Windhorst2018}
{Windhorst}, R.~A., {Timmes}, F.~X., {Wyithe}, J.~S.~B., {et~al.} 2018, \apjs,
  234, 41, \dodoi{10.3847/1538-4365/aaa760}

\bibitem[{{Zhang} {et~al.}(2015){Zhang}, {Yu}, \& {Lu}}]{Zhang2015}
{Zhang}, C., {Yu}, Q., \& {Lu}, Y. 2015, \apj, 813, 129,
  \dodoi{10.1088/0004-637X/813/2/129}

\bibitem[{{Zhang} {et~al.}(2018){Zhang}, {Yu}, \& {Lu}}]{Zhang2018}
---. 2018, \apj, 855, 36, \dodoi{10.3847/1538-4357/aaab4c}

\bibitem[{{Zitrin} {et~al.}(2013){Zitrin}, {Menanteau}, {Hughes}, {Coe},
  {Barrientos}, {Infante}, \& {Mandelbaum}}]{Zitrin2013}
{Zitrin}, A., {Menanteau}, F., {Hughes}, J.~P., {et~al.} 2013, \apjl, 770, L15,
  \dodoi{10.1088/2041-8205/770/1/L15}

\end{thebibliography}

%\bibliography{MyBiblio}{}
%\bibliographystyle{aasjournal}

\end{document}